\documentclass{jfm_arxiv}
\usepackage[applemac]{inputenc}
\usepackage{graphicx}
\usepackage{amsmath}
\usepackage{times}
\usepackage{txfonts}
\usepackage{array}
\usepackage{natbib}
\usepackage{rotating}
\usepackage{setspace}
\usepackage[dvipsnames,usenames]{color}

\newcommand{\te}[1]{\times10^{#1}}
\newcommand{\ut}[1]{\hspace{1mm}\mathrm{#1}}
\newcommand{\vc}[1]{\boldsymbol{\mathbi #1}}
\newcommand{\vecu}{\vc {u}} 
\newcommand{\vecB}{\vc {B}}

\newcommand{\ddp}[2]{\displaystyle \dfrac{\displaystyle \partial #1}{\displaystyle \partial #2}}

\newcommand{\parallelsum}{\mathbin{\!/\mkern-4mu/\!}}
\newcommand{\cit}[1]{\let\temp=\\ \centering #1\let\\=\temp}
\newcommand{\rev}[1]{{#1}}
\newcommand{\revv}[1]{{#1}}

\newcolumntype{L}[1]{>{\raggedright\let\newline\\\arraybackslash\hspace{0pt}}m{#1}}
\newcolumntype{C}[1]{>{\centering\let\newline\\\arraybackslash\hspace{0pt}}m{#1}}
\newcolumntype{R}[1]{>{\raggedleft\let\newline\\\arraybackslash\hspace{0pt}}m{#1}}

\shorttitle{Asymptotic convective dynamos}
\shortauthor{J. Aubert, T. Gastine, A. Fournier}

\title{Spherical convective dynamos in the rapidly rotating asymptotic regime}

\author{Julien Aubert
  \corresp{\email{aubert@ipgp.fr}},
  Thomas Gastine
 \and Alexandre Fournier}

\affiliation{Institut de Physique du Globe de Paris, Sorbonne Paris Cit\'e, Universit\'e Paris-Diderot, CNRS, 1 rue Jussieu, F-75005 Paris, France.}

\begin{document}

\maketitle

\begin{abstract}
Self-sustained convective dynamos in planetary systems operate in an asymptotic regime of rapid rotation, where a balance is thought to hold between the Coriolis, pressure, buoyancy and Lorentz forces (the MAC balance). Classical numerical solutions have previously been obtained in a regime of moderate rotation where viscous and inertial forces are still significant. We define a unidimensional path in parameter space between classical models and asymptotic conditions from the requirements to enforce a MAC balance and to preserve the ratio between the magnetic diffusion and convective overturn times (the magnetic Reynolds number). Direct numerical simulations performed along this path show that the spatial structure of the solution at scales larger than the magnetic dissipation length is largely invariant. This enables the definition of large-eddy simulations resting on the assumption that small-scale details of the hydrodynamic turbulence are irrelevant to the determination of the large-scale asymptotic state. These simulations are shown to be in \rev{good}  agreement with direct simulations in the range where both are feasible, and can be computed for control parameter values far beyond the current state of the art, such as an Ekman number $E=10^{-8}$. We obtain strong-field convective dynamos approaching the MAC balance and a Taylor state to an unprecedented degree of accuracy. The physical connection between classical models and asymptotic conditions is shown to be devoid of abrupt transitions, demonstrating the asymptotic relevance of classical numerical dynamo mechanisms. The fields of the system are confirmed to follow diffusivity-free, power-based scaling laws along the path.
\end{abstract}

\begin{keywords}
Dynamo theory, Magnetohydrodynamics, Geodynamo
\end{keywords}

\section{\label{intro}Introduction}
Convective dynamos in electrically conducting planetary fluid cores operate in an asymptotic limit of rapid rotation and strong energy input. Considering the Earth's core as representative of this regime, as we will do throughout this work, the values of a few dimensionless numbers may illustrate the situation (see table \ref{numbers}). Using a planetary rotation rate $\Omega=7.29\te{-5} \ut{s^{-1}}$, viscous and magnetic diffusivities $\nu\approx 10^{-6} \ut{m^{2}/s}$, $\eta\approx 1 \ut{m^{2}/s}$, a typical length scale $D=2260 \ut{km}$ and a typical velocity $U=0.5 \ut{mm/s}$, the Ekman number representing the ratio of the rotation period to the viscous diffusion time is $E=\nu/\Omega D^{2}=\mathcal{O}(10^{-15})$, implying a nearly inviscid behavior. The hydrodynamic Reynolds number measuring the ratio of the viscous diffusion time to the convective overturn time is $Re=UD/\nu=\mathcal{O}(10^{9})$, implying a developed hydrodynamic turbulence that is also strongly constrained by rotation, as witnessed by the low Rossby number $Ro=E Re = \mathcal{O}(10^{-6})$. The magnetic diffusion time relative to the convective overturn time yields the magnetic Reynolds number $Rm=UD/\eta=\mathcal{O}(10^{3})$, implying that the level of magnetic turbulence is more modest, though. The large disparity between the magnetic and viscous diffusion times, expressed by the small values of the magnetic Prandtl number $Pm=\nu/\eta=\mathcal{O}(10^{-{6}})$, implies a dominant ohmic dissipation of the injected power, and also presumably a high level of scale separation between the small-scale velocity field and a large-scale, self-sustained magnetic field, a situation which is classically intractable in the framework of global numerical simulation. 

The activity of numerical dynamo modelling has nevertheless blossomed in the space defined by modest, numerically tractable values of the aforementioned parameters (typically $E\ge10^{-5}$ and $Pm\ge0.2$ for full domain calculations) with interesting success in accounting for the static and kinematic morphological properties of Earth's magnetic field \citep[e.g.][]{Christensen2010, Aubert2013b}, a result that may be considered surprising given the distance in parameter space between such models and planetary conditions. The exploration of this classical parameter space has led to the formulation of scaling principles for the amplitude of the relevant fields of the system \citep{ChristensenAubert2006, Yadav2013FS,Yadav2013}. The power-based, rotation- and diffusivity-independent magnetic field scaling law proposed by these authors is now accepted given its success in accounting for a wide variety of planetary and stellar objects \citep{Christensen2009}. However, the theory underlying the flow speed, length scale and \rev{heat transfer} scalings has in general been subject to considerable debate, as it has been shown that the data set acquired in the classical parameter space can support multiple interpretations \citep{Soderlund2012,King2013,Davidson2013,Oruba2014}. Within this parameter space indeed, the part of the Coriolis force not balanced by the pressure gradient can be equilibrated by several different combinations of the other remaining forces (Lorentz force, buoyancy force, inertia, viscous forces) because none of these is really negligible. Another problem is that scaling predictions from the concurrent theories that have been formulated yield comparable levels of variance reduction against the numerical data set, and the scaling exponents are often too close to each other to be straightfowardly discriminable over the limited available control parameter range, unless one resorts to advanced statistical methods \citep{Stelzer2013}. Finally, the possible artificial alignment of data due to the choice of diffusivity-free scaling parameters has also been questioned \citep{Aurnou2007,Cheng2015}. 

These results lead to the conclusion that further insight on the planetary regime may be difficult to obtain without adding significantly more extreme numerical calculations to the original data set, a difficult task that in the past decade has received less attention than reanalyses of already available data. Still, a small number of extreme runs have been carried out at typical parameters $E\ge10^{-7}$ and $Pm\ge0.05$ \citep{Kageyama2008,Sakuraba2009,Miyagoshi2010,Sheyko2016,Nataf2015}, exhibiting a suprising variety of behaviors. In his comment of the study by \cite{Kageyama2008}, \cite{Christensen2008NV} underlines the usefulness of such runs to test and challenge our prior understanding, but points out that these can be fully rationalised only if they preserve the dynamical equilibria that are already well simulated, while progressively enforcing those that are not yet well accounted for. The difficulty resides in defining a corresponding path in parameter space between classical models and asymptotic conditions, which forms the first goal of this work. 

From a theoretical standpoint, \rev{a MAC (Magneto-Archimedes-Coriolis, also sometimes referred to as magnetostrophic) force balance should be enforced in Earth's core, or asymptotic} conditions \citep[e.g.][]{Braginsky1967,Starchenko2002,Davidson2013}. \revv{In the resulting dynamics, the} magnetic and buoyancy forces equilibrate the part of the Coriolis force not balanced by the pressure gradient, because of the relative smallness of viscous and inertial forces (small Ekman and Rossby numbers). From this follows the Taylor constraint \citep{Taylor1963} stating that the integral of the azimuthal magnetic force over cylinders co-axial with the rotation axis (the axial cylinders) should vanish in the limit $E, Ro \rightarrow 0$. Indeed, the pressure force identically vanishes on axial cylinders, as does the azimuthal Coriolis force, and buoyancy does not have an azimuthal component. In a situation where the magnetic force equilibrates with the Coriolis force while strongly dominating the fluid inertia, the magnetic energy should also largely dominate the kinetic energy \citep[e.g.][]{Davidson2013}, a regime which is commonly referred to as strong-field dynamo action. In the parameter space sampled so far by numerical simulations and laboratory experiments, a number of \rev{indirect diagnostics} are suggestive of the \rev{emergence of an} asymptotic regime: change in the dominant length scale of convection \citep{Sakuraba2009,Takahashi2012,Hughes2016}, in the efficiency of heat transfer \citep{King2015,Yadav2016}, and local cancellation between the Coriolis and Lorentz forces \citep[e.g.][]{Dormy2016}. Yet, characterising the asymptotic regime, and clearly separating the MAC forces from the residual contributions of inertia and viscosity definitely require to obtain models operating at control parameter values significantly outside the \rev{currently accessible space}. \rev{Likewise,} the MAC balance, the associated magnetostrophic Taylor state and strong-field dynamo action \rev{are more convincingly assessed by examining direct diagnostics such as} the actual levels of all forces \citep[e.g.][]{Wicht2010,Soderlund2015,Yadav2016PNAS}, the level of Taylor constraint enforcement \citep[e.g.][]{Wicht2010,Teed2015}, and the ratio of kinetic to magnetic energy \citep[e.g.][]{Takahashi2012}. \rev{These last two points} form the second goal of this work.

The key to achieve both goals is to introduce a relevant path in parameter space. It has been recognized \citep{Christensen2010}  that morphological semblance of the numerical dynamo output to the geomagnetic field can be achieved if a few time scale ratios are either set to, or brought reasonably close to\revv{,} their Earth counterparts. Among these, the Earth's magnetic Reynolds number value $Rm=\mathcal{O}(10^{3})$ is already numerically tractable. A sensible path connecting the available numerical models and the Earth should thus at least preserve $Rm$ while bringing other less-well simulated time scale ratios such as the Ekman and Rossby numbers to more Earth-like values (again, by reaching conditions of rapid rotation). Here, we show that the requirements to preserve $Rm$ and to observe the MAC balance enable the formulation of a unidimensional path between the currently available models and the Earth, such that all control parameters are determined by powers of a single path parameter $\epsilon$. In the formulation to be detailed in section \ref{path}, the value $\epsilon=1$ characterises the classical moderate models, while the Earth's core conditions correspond to $\epsilon=10^{-7}$, a value representative of the rapidly rotating limit $\epsilon \rightarrow 0$. The idea of a unidimensional path is equivalent to the mathematical concept of a distinguished limit, the relevance of which has also recently been suggested by \cite{Dormy2016} along somewhat different lines of reasoning.

As we will show, direct numerical simulation currently makes only a small portion of this path accessible, down to $\epsilon=0.1$. Within this subset, the large-scale model output shows a large degree of spatial invariance as $\epsilon$ is decreased. Large-scale invariance suggests that one may parameterise the effects of hydrodynamic turbulence below the magnetic dissipation length scale. This idea is in fact reminiscent of the large-eddy numerical simulations that have been exploited at the historical outset of the discipline \citep{Glatzmaier1995} for reasons of computational \rev{limitations}. Simple functional forms of hyperdiffusivity \citep[e.g.][]{Zhang1997,Grote2000} however may introduce dramatic changes in the character of the solution, though it is safe to use them at the end of the spectrum to stabilise the computation \cite[e.g.][]{Gastine2014}. Sophisticated subgrid parameterisations \citep[e.g.][]{Baerenzung2008,Baerenzung2010,Matsui2013} are promising but have never been fully tested in the spherical, self-sustained configuration examined in this work. Our approach combines physical arguments relative to \rev{the scales for energy injection and dissipation in the system}, together with a hyperdiffusive treatment recently proposed by \cite{Nataf2015}, applied to the velocity and density anomaly fields, \rev{but not to the magnetic field which} remains fully resolved. We obtain a much more tractable, but \rev{still reasonably} accurate large-eddy simulation, enabling the computation of models down to $\epsilon=3\te{-4}$, halfway in logarithmic scale between the moderate models and Earth along the parameter space path. This corresponds to the lowest Ekman number $E=10^{-8}$ reached to date in three-dimensional numerical dynamo simulation, to a low $Pm=0.045$, and more generally to diagnostic dimensionless numbers significantly more realistic than in the classical parameter space (table \ref{numbers}). \rev{Such models enable an exploration of the asymptotic behavior of spherical convective dynamos. This study is organised as follows:} section \ref{model} describes the numerical model and methods. The path theory and numerical results are presented in section \ref{path} and \ref{results}, and are discussed in section \ref{discu}.

\section{\label{model}Numerical model}
\subsection{Model equations, input parameters and methods}

We consider a convecting, electrically conducting, incompressible fluid in a self-gravitating spherical shell between radii $r_{i}$ and $r_{o}$, with $r_{i}/r_{o}=0.35$ as in the Earth's core at present. The shell is rotating about an axis $\vc{e}_z$ with an angular velocity $\Omega$. We solve for Boussinesq convection for a velocity field $\vc{u}$ and a density anomaly field $C$, magnetic induction in the magnetohydrodynamic approximation for a magnetic field $\vc{B}$, with the following set of dimensionless equations:
\begin{gather}
\ddp{\vecu}{t}+\vecu \cdot \nabla\vecu+2~\vc{e}_z\times\vecu+\nabla P=Ra^{*}_{(F)}~\dfrac{\vc{r}}{r_o} C + \left(\nabla\times\vecB\right)\times\vecB+E\nabla^2\vecu\label{NS},\\
\ddp{\vecB}{t}=\nabla\times\left(\vecu\times\vecB\right)+\dfrac{E}{Pm}\nabla^2\vecB\label{MI},\\
\ddp{C}{t}+\vecu \cdot \nabla C= \dfrac{E}{Pr} \nabla^2 C+S \label{thermochem},\\
\nabla\cdot\vecu=0,\label{solu}\\
\nabla\cdot\vecB=0.\label{solb}
\end{gather}

\begin{table}
\begin{center}
\setstretch{1}
\begin{tabular}{lccccl}
Name & Symbol & Definition & Meaning & Range in this work & Earth\\[3pt]
Ekman & $E$ & $\nu/\Omega D^{2}$ & $\dfrac{\text{planetary rotation period}}{\text{viscous diffusion time}}$ & $10^{-8}-3\te{-5}$ &$\mathcal{O}(10^{-15})$\\[10pt]
Magnetic Prandtl & $Pm$ & $\nu/\rev{\eta}$ & $\dfrac{\text{magnetic diffusion time}}{\text{viscous diffusion time}}$ & $0.045 - 2.5$ & $\mathcal{O}(10^{-6})$ \\[10pt]
Flux Rayleigh & $Ra^{*}_{F}$ & $\dfrac{g_{o}F}{4\pi\rho\Omega^3 D^4}$  & $\left(\dfrac{\text{power input rate}}{\text{planetary rotation rate}}\right)^{3}$ & $9\te{-9}-2.9\te{-5}$ & $\mathcal{O}(10^{-12})$\\[10pt]
Lehnert & $\lambda$ & $\dfrac{B}{(\rho\mu)^{1/2}\Omega D}$ & $\dfrac{\text{planetary rotation period}}{\text{Alfv\'en time}}$ & $2\te{-3}-2\te{-2}$ & $\mathcal{O}(10^{-4})$ \\[10pt]
Elsasser & $\Lambda$ & $B^{2}/\rho \mu \eta \Omega$ & $\dfrac{\text{Lorentz force}}{\text{Coriolis force}}$ & $18.4-31.7$ & $\mathcal{O}(10)$ \\[10pt]
Rossby & $Ro$ & $U/\Omega D$ & $\dfrac{\text{planetary rotation period}}{\text{overturn time}}$ & $2.4\te{-4}-1.1\te{-2}$  & $\mathcal{O}(10^{-6})$   \\[10pt]
Reynolds & $Re$ & $U D / \nu$  & $\dfrac{\text{viscous diffusion time}}{\text{overturn time}}$ & $372-2.4\te{4}$  & $\mathcal{O}(10^{9})$ \\[10pt]
Magnetic Reynolds & $Rm$ & $U D / \eta$  & $\dfrac{\text{magnetic diffusion time}}{\text{overturn time}}$ & $930-1099$  & $\mathcal{O}(10^{3})$ \\[10pt]
squared Alfv\'en & $A^{2}$ & $\rho\mu U^{2}/B^{2}$ & $\dfrac{\text{kinetic energy}}{\text{magnetic energy}}$ & $1.4\te{-2}-0.48$ & $\mathcal{O}(10^{-4})$ \\
\end{tabular}
\end{center}
\caption{\label{numbers}Summary of the main input and output parameters, together with their values in this work and in the Earth's core. See section \ref{model} for definitions and section \ref{path} for a discussion of geophysical estimates.}
\end{table}

Here $\vc{r}$ is the radius vector. Time is scaled with the inverse of the rotation rate $\Omega^{-1}$. Length is scaled with the shell gap $D=r_{o}-r_{i}$. Velocity is scaled with $\Omega D$.  Magnetic induction is scaled by $(\rho \mu)^{1/2} \Omega D$, where $\rho$ is the fluid density and $\mu$ the magnetic permeability of the fluid. Tables \ref{numbers} and \ref{table} summarise the values of the input parameters for our survey, together with their Earth estimates. We recall the definitions for the Ekman number $E$, magnetic Prandtl number $Pm$, and introduce the Prandtl number $Pr$ as:
\begin{gather}
E=\dfrac{\nu}{\Omega D^2},\\
Pm=\dfrac{\nu}{\eta},\\
Pr=\dfrac{\nu}{\kappa}.
\end{gather}
Here $\nu,\rev{\eta}$ have already been introduced as the fluid viscous of magnetic diffusivities, and $\kappa$ is the diffusivity of the density anomaly field $C$. Two different sets of boundary conditions are investigated. Type ST refers to the standard boundary conditions \rev{often} used in the numerical dynamo modelling community \citep[e.g.][]{ChristensenAubert2006}: rigid (no-slip) and electrically insulating boundary conditions, and fixed density anomaly at both boundaries, with an imposed difference $\Delta C$ between inner and outer boundary (which also serves as the scale for the dimensionless density anomaly). In this case the density anomaly source term $S$ vanishes in equation (\ref{thermochem}), and the modified Rayleigh number $Ra^*$ appearing in equation (\ref{NS}) \rev{reads}:
\begin{equation}
Ra^*=\dfrac{g_{o} \Delta C}{\rho \Omega^{2} D}.
\end{equation}
Here $g_{o}$ is the gravity at the outer boundary. The modified Rayleigh number relates to the canonical Rayleigh number $Ra=g_{o}\Delta C D^{3}/ \rho \kappa\nu$ through
\begin{equation}
Ra^*=Ra E^{2} /Pr.\label{cantomod}
\end{equation}
Note that the square root of the modified Rayleigh number is classically referred to as the convective Rossby number \citep{Gilman1977}. The second set of boundary conditions (type CE) is derived from the Coupled Earth dynamo setup \citep{Aubert2013b}. Here the idea is to minimise the residual influence of boundary layers, and provide a geophysically more relevant description of global coupling by using stress-free, electrically conducting, and fixed mass anomaly flux conditions at both boundaries. We will show in section \ref{results} that results obtained within setups ST and CE are largely similar, meaning that the nature of the asymptotic state is not influenced by the choice of boundary conditions. In type CE, the imposed mass anomaly flux crossing the shell is defined as $F$, and the density anomaly scale for nondimensionalisation is defined as $F/4\pi \Omega D^{3}$.  The dimensionless, spherical parts of the mass anomaly flux at the inner and outer boundaries are fixed to $\partial C / \partial r (r_{i})= -Pr/\,E r_{i}^{2}$ and $\partial C / \partial r (r_{o})=0$. The modified flux Rayleigh number appearing in equation (\ref{NS}) then \rev{becomes}
\begin{equation}
Ra^{*}_{F}=\dfrac{g_{o}F}{4\pi\rho\Omega^3 D^4},\label{fluxra}
\end{equation}
and mass flux conservation implies that the source term in equation (\ref{thermochem}) for type CE \rev{is} $S=-3/(r_{o}^{3}-r_{i}^{3})$ \citep{Aubert2009}. Throughout this work, the notation $Ra^{*}_{(F)}$ introduced in equation (\ref{NS}) refers either to $Ra^{*}$ or $Ra^{*}_{F}$ depending on whether we consider the ST or the CE setup, respectively.

In addition to this basic setup, type CE also features a number of refinements, listed below, which help the simulation output to match the statics and kinematics of the geomagnetic field, particularly regarding the geomagnetic westward drift and the geographical localisation of the temporal magnetic field variations. Although these are mostly irrelevant to the fundamental force balance and scaling results to be presented in the following sections, they have been kept here for future exploitation of the results in a geophysical context. The fluid shell is magnetically coupled to a solid inner sphere of radius $r_{i}$ \rev{(the inner core)} and with the same electrical conductivity, axially rotating at a rotation rate $\Omega_{ic}$. The inner sphere couples to a solid spherical shell between radii $r_{o}$ and $1.83r_{o}$ \rev{(the mantle)}, also axially rotating at a rate $\Omega_{m}$, through a restoring remote torque $\Gamma=\xi (\Omega_{ic}-\Omega_{m})$ \rev{meant to arise from gravitational coupling}, with $\xi=0.75$. The outer solid shell has an dimensionless electrical conductance $\Sigma=10^{-4}$ also enabling magnetic coupling with the fluid shell \citep[see][eqns. 28 and 29, for complete definitions of $\xi$ and $\Sigma$]{Pichon2016}.  Angular momentum of the coupled system is preserved, with moments of inertia for the inner sphere, fluid shell, and outer solid shell in Earth-like ratios \citep[see also][]{Pichon2016}. Finally, a longitudinal hemispheric modulation of the mass anomaly flux is imposed at the inner core boundary on top of the homogeneous part, with a peak-to-peak amplitude $0.8 Pr/\,Er_{i}^{2}$. A spatial modulation of the mass anomaly flux at the outer boundary is also imposed, with a peak-to-peak amplitude $0.1 Pr/\,E$. The patterns and relative orientations of these mass anomaly flux heterogeneities are those prescribed in \cite{Aubert2013b}.

We have carried out 10 numerical cases of type CE and 7 cases of type ST (table \ref{table}). Type ST cases were integrated using the MagIC numerical implementation \citep[][freely available at www.github.com/magic-sph/magic]{Wicht2002,MagIC}, which uses a Chebyshev decomposition in the radial direction with $NR$ nodal points, and a lateral spherical harmonic decomposition up to degree and order $l_{\text{max}}$. Type CE cases were integrated using the PARODY-JA numerical implementation \citep{Dormy1998,Aubert2008} using a second-order finite differencing scheme in the radial direction with $NR$ grid points in the fluid, 50 grid points in the inner sphere, and the same spherical harmonic decomposition as MagIC. Both implementations use a toroidal-poloidal decomposition of the solenoidal vector fields $\vecu$ and $\vecB$ (equations \ref{solu},\ref{solb}), the same Crank-Nicolson/Adams-Bashforth semi-implicit adaptative time stepping scheme of second order in time, and are benchmarked against each other \citep{Christ2001}. Both implementations resort to the library SHTns \citep[][freely available at https://bitbucket.org/nschaeff/shtns]{Schaeffer2013}, for efficient handling of the spherical harmonic transforms. All calculations have been performed in the full domain and no azimuthal symmetry was assumed. More extreme parameter regimes could be reached in the CE setup, owing to the absence of viscous and density anomaly boundary layers. The least forced reference models 0 and 1 in the ST and CE setups were already largely supercritical, with Rayleigh numbers respectively 40 and 400 times their critical values. Note however that these figures are different because supercriticality is defined relative to the flux in CE case, and the dynamical regimes of models 0 and 1 are otherwise equivalent. All models produced a self-sustained magnetic field with a dominant axial dipole that did not reverse polarity, and with a morphology exhibiting detailed resemblance to that of the geomagnetic field (see section \ref{res:structure}). No bistable states were found, though such states were not specifically searched for. \rev{Our initial conditions for models 0 and 1 were indeed an axially dipolar magnetic field, zero velocity field and a random density anomaly perturbation. We then progressed by half-decades towards lower Ekman numbers by iteratively initialising each new model with the final state of the previous calculation. Large-eddy simulations (section \ref{LES}) were computed first, and then expanded into fully resolved numerical simulations by removing the hyperdiffusive treatment, a strategy causing only weak transients and ensuring significant savings in numerical cost.}

\begin{sidewaystable}
\setstretch{1}
\vspace*{15cm}\begin{tabular}{cllllrrrrrcrcccl}
Label &Type & $NR$ & $l_{\text{max}}$ & $q_{h}$ & $\epsilon$ & $E$ & $Pm$ & $Ra^{*}_{(F)}$ & $Nu$ & $p$ & $Rm$ & $\Lambda$ & $f_{\text{ohm}}$ & $d_{\text{min}}$ &  $\mathcal{T}$ \\[3pt]
Model 0& CE/D & 200 & 133 &             & $ 1 $                  & $ 3\te{-5} $ & $ 2.5 $    & $ 2.7\te{-5} $  &                  & $ 8.56\te{-6} $ & $ 930 $   & $ 21.4 $ & $ 0.60 $ & $ 2.44\te{-2} $ &  $ 0.27 $ \\
& CE/D & 260 & 170 &             & $ 0.33 $             & $ 10^{-5} $ & $ 1.44 $  & $ 9\te{-6} $     &                  & $ 2.93\te{-6} $ & $ 1036 $ & $ 20.3 $ & $ 0.71 $ & $ 2.17\te{-2} $ &  $ 0.22 $ \\
& CE/D & 480 & 256 &             & $ 0.1 $               & $ 3\te{-6} $ & $ 0.8 $    & $ 2.7\te{-6} $  &                  & $ 8.96\te{-7} $ & $ 1092 $ & $ 20.4 $ & $ 0.81 $ & $ 1.98\te{-2} $ &  $ 0.17 $ \\
& CE/L  & 180 & 133 & 1.0325 & $ 0.33 $             & $ 10^{-5} $ & $ 1.44 $  & $ 9\te{-6} $     &                  & $ 2.94\te{-6} $ & $ 988 $   & $ 20.7 $ & $ 0.67 $ & $ 2.25\te{-2} $ &  $ 0.22 $ \\
& CE/L  & 200 & 133 & 1.045   & $ 0.1 $               & $ 3\te{-6} $ & $ 0.8 $    & $ 2.7\te{-6} $  &                  & $ 9.08\te{-7} $ & $ 1046 $ & $ 21.6 $ & $ 0.72 $ & $ 2.15\te{-2} $ & $ 0.17 $ \\
& CE/L  & 240 & 133 & 1.0575 & $ 3.33\te{-2} $ & $ 10^{-6} $ & $ 0.45 $  & $ 9\te{-7} $     &                  & $ 3.09\te{-7} $ & $ 1036 $ & $ 20.9 $ & $ 0.76 $ & $ 2.09\te{-2} $ & $0.12 $ \\
& CE/L  & 320 & 133 & 1.07     & $ 0.01 $             & $ 3\te{-7} $ & $ 0.25 $  & $ 2.7\te{-7} $  &                  & $ 9.39\te{-8} $ & $ 1046 $ & $ 22.1 $ & $ 0.80 $ & $ 2.06\te{-2} $ & $0.082$ \\
& CE/L  & 400 & 133 & 1.082   & $ 3.33\te{-3} $ & $ 10^{-7} $ & $ 0.144 $ & $ 9\te{-8} $    &                  & $ 3.17\te{-8} $ & $ 1068 $ & $ 21.9 $ & $ 0.81 $ & $ 2.03\te{-2} $ & $0.058$ \\
& CE/L  & 504 & 133 & 1.11     & $ 10^{-3} $          & $ 3\te{-8} $ & $ 0.079 $ & $ 2.7\te{-8} $ &                  & $ 9.62\te{-9} $ & $ 1099 $ & $ 20.9 $ & $ 0.83 $ & $ 1.94\te{-2} $ & $ 0.045 $ \\
& CE/L  & 624 & 133 & 1.14     & $ 3.33\te{-4} $ & $ 10^{-8} $ & $ 0.045 $ & $ 9\te{-9} $    &                  & $ \rev{3.24\te{-9}} $ & $ \rev{1082} $ & $ \rev{19.9} $ & $ \rev{0.84} $ & $ \rev{1.90\te{-2}} $ &  $ 0.036 $ \\
Model 1& ST/D &   97 & 133 &             & $ 1 $                  & $ 3\te{-5} $ & $ 2.5 $    & $ 0.11 $           & $ 9.9 $    & $ 1.44\te{-5} $ & $ 934 $   & $ 30.6 $ & $ 0.57 $ & $ 2.31\te{-2} $ \\
& ST/D & 121 & 256 &             & $ 0.34 $             & $ 10^{-5} $ & $ 1.44 $  & $ 0.07 $            & $ 15.4 $ & $ 4.87\te{-6} $ & $ 970 $   & $ 31.0 $ & $ 0.70 $ & $ 2.09\te{-2} $ & \\
& ST/D & 217 & 341 &             & $ 0.10 $             & $ 3\te{-6} $ & $ 0.79 $  & $ 4.21\te{-2} $& $ 25.5 $ & $ 1.51\te{-6} $ & $ 1055 $ & $ 29.4 $ & $ 0.80 $ & $ 1.88\te{-2} $ &  \\
& ST/L  & 121 & 170 & 1.0325  & $ 0.32 $             & $ 10^{-5} $ & $ 1.44 $  & $ 0.07 $            & $ 14.8 $ & $ 4.66\te{-6} $ & $ 953 $   & $ 31.7 $ & $ 0.68 $ & $ 2.19\te{-2} $ &  \\
& ST/L  & 193 & 170 & 1.045    & $ 8.91\te{-2} $ & $ 3\te{-6} $ & $ 0.79 $  & $ 4.21\te{-2} $ & $ 22.1 $ & $ 1.29\te{-6} $ & $ 982 $   & $ 29.6 $ & $ 0.75 $ & $ 2.10\te{-2} $ &  \\
& ST/L  & 201 & 170 & 1.055    & $ 2.60\te{-2} $ & $ 10^{-6} $ & $ 0.46 $  & $ 2.66\te{-2} $ & $ 30.2 $ & $ 3.75\te{-7} $ & $ 937 $   & $ 27.8 $ & $ 0.79 $ & $ 2.13\te{-2} $ &  \\
& ST/L  & 289 & 133 & 1.07      & $ 6.42\te{-3} $ & $ 3\te{-7} $ & $ 0.25 $  & $ 1.60\te{-2} $ & $ 41.0 $ & $ 9.27\te{-8} $ & $ 907 $   & $ 24.6 $ & $ 0.80 $ & $ 2.18\te{-2} $ & \\
\\
Earth & &        &         &             &  $\mathcal{O}(10^{-7})$ & $\mathcal{O}(10^{-15})$ & $\mathcal{O}(10^{-6})$ & $\mathcal{O}(10^{-12})$  & & $\mathcal{O}(10^{-12})$ & $\mathcal{O}(10^{3})$ & $\mathcal{O}(10)$ & 1 & $\mathcal{O}(10^{-2})$ &  $\mathcal{O}(10^{-3})$ \\
& & & & & & & & & & & & & & & (this work)
  \end{tabular}
  \caption{\label{table}Set of numerical models with input and output parameters (see section \ref{model} for definitions), together with estimates for Earth's core (see section \ref{path}). Type CE refers to numerical models performed in the Coupled Earth setup (fixed mass anomaly flux, stress-free, electrically conducting boundary conditions). Type ST refers to models performed in the standard setup (fixed density anomaly, no-slip, electrically insulating boundary conditions). Types D and L respectively refer to direct and large-eddy simulations (see section \ref{LES}). Model 0 is similar to the original CE dynamo \citep[][parameters with subscript 0 in the text]{Aubert2013b}. Model 1 is an equivalent standard dynamo (parameters with subscript 1 in the text). All models have $r_{i}/r_{o}=0.35$ and $Pr=1$. All large-eddy simulations have $l_h=30$. Note that the magnetic Reynolds number $Rm$ is constant to within $\pm 10\%$, and that the Elsasser number $\Lambda$ for a given boundary condition type (CE or ST) is also constant to within $\pm 10\%$.}
\end{sidewaystable}

\subsection{Output parameters}
In the following, we will analyse the following time-averaged, integral outputs, which are also summarised in table \ref{numbers} and detailed in table \ref{table}. The magnetic field amplitude in the shell is characterised either by the Lehnert number 
\begin{equation}
\lambda= \dfrac{B}{(\rho\mu)^{1/2}\Omega D},
\end{equation} or the Elsasser number 
\begin{equation}
\Lambda=\dfrac{B^{2}}{\rho \mu \eta \Omega},
\end{equation}
where $B$ is the dimensional, root-mean-squared magnetic field amplitude in the fluid shell. Table \ref{table} lists the Elsasser numbers in our numerical cases, and we recall that $\lambda=\sqrt{\Lambda E/Pm}$. The velocity field amplitude is characterised either by the Rossby number 
\begin{equation}
Ro=\dfrac{U}{\Omega D},
\end{equation} 
the hydrodynamic Reynolds number 
\begin{equation}
Re=\dfrac{U D }{\nu},
\end{equation}
or the magnetic Reynolds number 
\begin{equation}
Rm=\dfrac{U D}{\eta},
\end{equation} 
where $U$ is the dimensional, root-mean-squared velocity field amplitude in the fluid shell. Table \ref{table} lists the magnetic Reynolds numbers, and we recall that $Re=Rm/Pm$ and $Ro=E Rm/Pm$. The squared Alfv\'en number 
\begin{equation}
A^{2}=\dfrac{\rho\mu U^{2}}{B^{2}},
\end{equation}
measuring the ratio of kinetic to magnetic energy can also be derived from table \ref{table} as $A^{2}=E Rm^2 / Pm \Lambda$. The efficiency of mass anomaly transfer in the ST cases is measured by the Nusselt number $Nu= FD/4\pi r_{i}r_{o}\kappa \Delta C$. 

\begin{figure}
\centerline{\includegraphics[width=13.5cm]{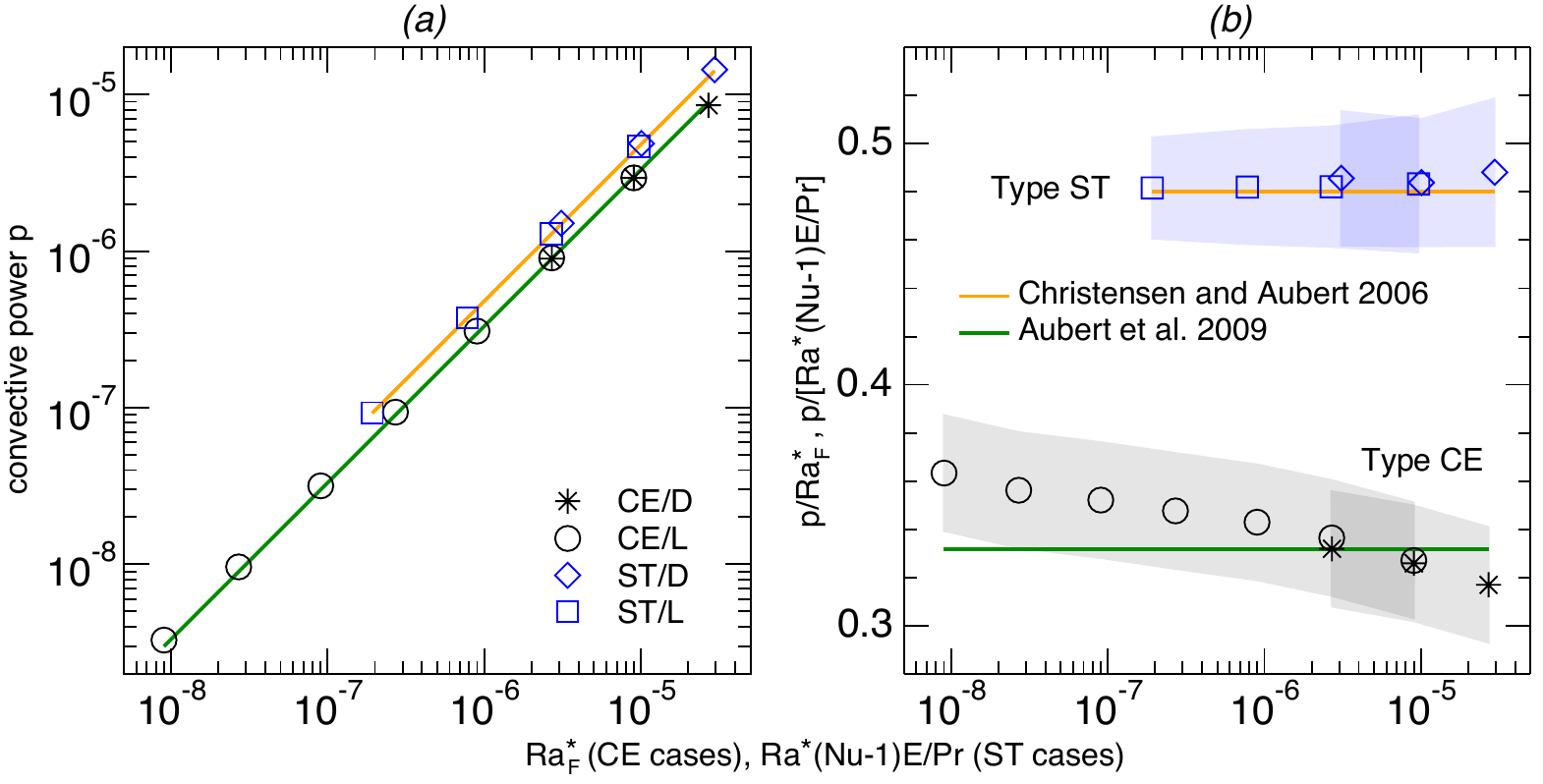}}
\caption{\label{Rap}Proportionality ({\it a}) between the convective power $p$ and the dimensionless mass anomaly flux $Ra^{*}_{F}$ (CE cases) or $Ra^{*}(Nu-1)E/Pr$ (ST cases), presented together with its compensated scaling ({\it b}). The green line is the analytic prediction relevant to the CE setup from the perfect-mixing theory \citep{Aubert2009}, with proportionality constant $\gamma=0.332$. The orange line is the analytic prediction $\gamma=0.480$ obtained in the ST setup by \cite{ChristensenAubert2006}. In panel ({\it b}), shaded regions represent the $\pm 1$ std. dev. of temporal fluctuations relative to the time average. In the CE setup, the large-eddy simulations (see section \ref{LES}) exhibit a weak residual trend for $\gamma$ which should amount to about $20 \%$ at Earth's core value $Ra^{*}_{F}=O(10^{-12})$, a negligible error compared to the geophysical uncertainty in the determination of $Ra^{*}_{F}$ (equation \ref{RaEarth}).}
\end{figure}

The volumetric convective power is defined as
\begin{equation}
p=\dfrac{Ra^{*}_{(F)}}{Vr_{o}} \int_{V} (\vc{u} \cdot \vc{r})C\,\mathrm{d}V,\label{pdef}
\end{equation}
where $V=4\pi(r_{o}^{3}-r_{i}^{3})/3$ is the volume of the fluid shell. On time average, the convective power equates \rev{to} the rate of gravitational potential energy release, which itself \rev{relates} to the mass anomaly flux. \rev{In the CE setup this leads to a relationship between $p$ and $Ra^{*}_{F}$ \citep{Aubert2009} while in the ST setup this may be more precisely assessed by relating $p$ to an advected flux} $Ra^{*}(Nu-1)E/Pr$ \citep{ChristensenAubert2006}. Determining the proportionality factor $\gamma=p/Ra^{*}_{F}$ (CE setup) or $\gamma=p/(Ra^{*}(Nu-1)E/Pr)$ (ST setup) requires knowledge of the gravitational potential difference between the radii for density anomaly injection and mixing. The constant $\gamma$ can be exactly determined in a configuration of condensed mass central gravity $g\sim 1/r^{2}$ \citep{Gastine2015}. Approximate values of $\gamma$ depending only on geometry can also be analytically determined in the present situation of sufficiently supercritical convection and linear radial gravity, through an assumption on the density anomaly profile \citep[ST setup,][]{ChristensenAubert2006} or on the average gravitational potential relevant to density anomaly mixing \citep[CE setup,][]{Aubert2009}. Figure \ref{Rap} shows that the deviations of our simulations from these approximate theories remain weak throughout the investigated range and should also remain negligible at Earth's core conditions\rev{, as expected for strongly supercritical convection in the rotationally-dominated regime \citep{Oruba2016}}. In the following, we will thus consider that there is proportionality between $p$ and $Ra^{*}_{F}$ in the CE setup, such that power is an input parameter. In the ST setup we will also consider that $p$ is proportional to $Ra^{*}(Nu-1)E/Pr$, meaning that power is an output parameter.

The volumetric rate of ohmic dissipation is defined as
\begin{equation}
D_{\eta}=\dfrac{E}{PmV} \int_{V} (\nabla\times\vecB)^{2}\,\mathrm{d}V,
\end{equation}
and is expressed in table \ref{table} through the ohmic dissipation fraction $f_{\text{ohm}}=D_{\eta}/p$. The equivalent ohmic dissipation length as defined in \cite{Christensen2004} is also reported in table \ref{table} and \rev{reads}
\begin{equation}
d_{\text{min}}=\rev{\left(\int_{V} \vecB^{2}\,\mathrm{d}V\right)^{1/2}\bigg/\left(\int_{V} (\nabla\times\vecB)^{2}\,\mathrm{d}V\right)^{1/2}}= \dfrac{E}{Pm} \sqrt{\dfrac{\Lambda}{p f_{\text{ohm}}}}.
\end{equation}

We finally use a standard measure \citep{Wicht2010,Teed2015} for the enforcement of the Taylor constraint \citep{Taylor1963}, by assessing the cancellation level of the Lorentz force acting on axial cylinders:
\begin{equation}
\mathcal{T}(s)=\dfrac{ \displaystyle\int_{z_{-}}^{z_{+}} \vc{e}_{\varphi}\cdot\overline{\left(\nabla\times\vc{B}\right)\times\vc{B}}\,\mathrm{d}z}{\displaystyle\int_{z_{-}}^{z_{+}} \left|\,\vc{e}_{\varphi}\cdot\overline{\left(\nabla\times\vc{B}\right)\times\vc{B}}\,\right|\,\mathrm{d}z}.\label{tayloreq}
\end{equation}
Here $s,\varphi,z$ are cylindrical coordinates, $\vc{e}_{\varphi}$ is the unit vector in the azimuthal direction, and the overbar represents an average taken in the azimuthal direction. The vertical integrals are evaluated between the lower and upper heights $z_{-,+}$ at which the axial cylinder of radius $s$ intersects the spherical shell boundaries. In table \ref{table} we report $\mathcal{T}$ as an average of $|\mathcal{T}(s)|$ over axial cylinders between $s=0$ and $s=r_{o}$, hence also encompassing fluid regions above and below the inner sphere inside the axial cylinder tangent to the inner core (the tangent cylinder).

\section{\label{path}A parameter space path connecting models to asymptotic conditions}

\rev{Here we first recall the MAC balance theory introduced by \cite{Davidson2013} (from hereafter D13). We use this theory as a guideline for defining a parameter space path connecting classical models to asymptotic conditions, along which the solutions should show a degree of large-scale invariance. From this we introduce large-eddy simulations aiming at capturing the essential features of the system at a drastically reduced computer cost. A spatially invariant, approximated version of the D13 theory is introduced to account for the scaling properties of these large-eddy simulations. The two theories are assessed in section \ref{results} against numerical data, and their agreement together with the validity of the associated approximations are discussed in section \ref{discu}.}

\subsection{\label{MACtheory}Outline of the MAC theory}
\rev{The D13 theory} introduces three length scales to describe the asymptotic solutions. \rev{First,} $d_{\parallelsum}$ describes the typical size of convection structure along the rotation axis. \rev{The second scale $d_{\perp}$ is} the typical size of convection structures in a plane perpendicular to the rotation axis. \rev{The third scale is finally} the magnetic dissipation length $d_{\text{min}}$ which we introduced above in its dimensionless form. The columnar structure of convection rolls is a robust feature of spherical convective dynamos \citep{ChristensenAubert2006,Yadav2013}, prompting 
\begin{equation}
d_{\parallelsum}\sim D.
\end{equation}
The MAC balance evaluated from the curled, dimensional version of equation (\ref{NS}) \rev{leads to}
\begin{equation}
\dfrac{\rho\Omega U}{D} \sim \dfrac{g_{o} C}{d_{\perp}} \sim \dfrac{B^{2}}{\mu d_{\perp}^{2}},\label{MACbal}
\end{equation} 
where $C$ denotes a typical density anomaly, complementing the already introduced typical velocity and magnetic fields $U$ and $B$. Likewise, the dimensional balance between the rate of convective energy input and magnetic dissipation \citep{Christensen2004,ChristensenAubert2006} may now be written
\begin{equation}
\dfrac{\eta B^{2}}{\mu d_{\text{min}}^{2}} \sim f_{\text{ohm}} \dfrac{g_{o} F}{D^{2}},\label{Ebudget}
\end{equation}
where we have used the equivalence between convective power and mass anomaly flux (figure \ref{Rap}). Using $F\sim UCD^{2}$, and assuming a context of dominant magnetic dissipation i.e. $f_{\text{ohm}}\approx 1$, one sees that satisfying both the balance between buoyancy and magnetic forces in (\ref{MACbal}) and the energy budget (\ref{Ebudget}) implies that 
\begin{equation}
U/d_{\perp}\sim\eta/d_{\text{min}}^{2}.\label{vorteq}
\end{equation}
This \rev{implies} that the large-scale vorticity and the vorticity at the scale of magnetic dissipation are \rev{equivalent. This is an important result as it indicates that the turbulent energy transfers which are essential to the dynamics occur between the large scale $d_{\perp}$ at which the magnetic field sustains itself by drawing convective power (balance between buoyancy and Lorentz forces), and the small scale $d_{\min}$ at which this power is ohmically dissipated.} If one further requires the large- and small-scale vorticities to be independent on the system rotation rate and diffusivities, then from (\ref{Ebudget}) it follows that the magnetic field itself is independent on the rotation rate and diffusivities. Dimensional analysis finally yields the following scaling, corresponding to the initial proposal of \cite{ChristensenAubert2006}:
\begin{equation}
B \sim \sqrt{\rho\mu} (g_{o}F/\rho D)^{1/3}.
\end{equation}
One finally obtains the dimensionless power-driven, diffusivity-free scalings 
\begin{eqnarray}
Ro \sim (Ra^{*}_{F})^{4/9},\label{D13U}\\
\lambda\sim (Ra^{*}_{F})^{1/3},\label{D13B}
\end{eqnarray} 
together with the following relationships for the length scales
\begin{eqnarray}
d_{\text{min}}/D &\sim& Rm^{-1/2} Ro^{1/8},\label{dminscaling}\\
d_{\perp}/D&\sim& Ro^{1/4}. \label{dperpscaling}
\end{eqnarray}
In equation (\ref{dminscaling}) $d_{\text{min}}$ thus primarily scales with the magnetic Reynolds number, in accordance with the initial result of \cite{Christensen2004}. In the range between model 0 (table \ref{table}) where $Ro=10^{-2}$, and the conditions of Earth's core where $Ro\approx 3\te{-6}$ (see section \ref{CEpath}), the secondary dependence with the Rossby number $Ro^{1/8}$ should remain small, within a factor 3. Numerical dynamo surveys \citep{Christensen2010ISSI,Yadav2013FS,Stelzer2013} generally suggest an even weaker dependence of $d_{\text{min}}$ in $Ro$ with power-law exponents between 1/20 and 1/10. The discrepancy with D13 may be rooted in the variations of $f_{\text{ohm}}$ that are still significant in the numerical data set but not considered in the asymptotic MAC scaling theory. A reasonable approximation is thus to neglect the Rossby number dependence in (\ref{dminscaling}). Consistency with the vorticity equivalence (\ref{vorteq}) then also demands \rev{the stronger assumption} to neglect the Rossby number dependence in (\ref{dperpscaling}).

\subsection{\label{CEpath}Path theory in the CE setup}
The calculations listed in table \ref{table} do not represent a systematic sampling of the parameter space, but are rather chosen to follow a path connecting the classical numerical models such as the original CE dynamo \citep{Aubert2013b} to Earth's core conditions. To introduce this concept it is best to use a reasoning based on time scale ratios, with our ultimate goal being to bring five relevant time scales in Earth-like proportions. These are the inverse rotation rate $\tau_{\Omega}=1/\Omega$, the convective overturn time $\tau_{U}=D/U$, the Alfv\'en time $\tau_{A}=\sqrt{\rho\mu}D/B$, the time scale for convective power input $\tau_{F}=(4\pi\rho D^4/g_{o}F)^{1/3}$ (analogous to the Kelvin-Helmholtz time scale), and the magnetic diffusion time $\tau_{\eta}=D^{2}/\eta$. Less relevant to us are the viscous diffusion time $\tau_{\nu}=D^{2}/\nu$ and the density anomaly diffusion time $\tau_{\kappa}=D^{2}/\kappa$. For these latter times we do not target Earth-like ratios relative to the other times, but simply conditions corresponding to the magnetic diffusion time $\tau_{\eta}$ being much smaller than $\tau_{\nu}$ and $\tau_{\kappa}$, so that ohmic losses are the dominant source of dissipation.

We first illustrate the strategy with CE cases. Our starting point is a model similar to the original coupled Earth dynamo \citep{Aubert2013b}, which we label model 0 (see table \ref{table}). In the following, subscript 0 denotes the parameters relative to this model. Four relevant time scale ratios for this model are
\begin{eqnarray}
\left(\dfrac{\tau_{\Omega}}{\tau_{F}}\right)^{3}&=&(Ra^{*}_{F})_{0}=2.7\te{-5},\\
\dfrac{\tau_{\Omega}}{\tau_{\eta}}&=&E_{0}/Pm_{0}=1.2\te{-5},\\
\dfrac{\tau_{\eta}\tau_{\Omega}}{\tau_{A}^{2}}&=&\Lambda_{0}=21.4,\\
\dfrac{\tau_{\eta}}{\tau_{U}}&=&Rm_{0}=930.
\end{eqnarray}
Our goal corresponds to conditions in the Earth's core. To estimate these we use the values $B=4 \ut{mT}$ \citep{Gillet2010}, $U=5\te{-4}\ut{m/s}$ \citep[e.g.][]{Aubert2014}, $\rho=11000 \ut{kg/m^{3}}$, $D=2260\ut{km}$, $\Omega=7.29\te{-5}\ut{s^{-1}}$, which are known with some certainty. We further use the less certain values $\eta=0.3-3\ut{m^{2}/s}$ for the magnetic diffusivity, and $Q_{ad}=4-15 \ut{TW}$ for the core-mantle boundary adiabatic heat flow, the uncertainty being due to an ongoing debate on core thermal and electrical conductivities \citep[e.g.][]{Pozzo2012,Ohta2016,Konopkova2016}. We assume that heat flow at the core-mantle boundary is exactly adiabatic, such that the dynamo is entirely bottom-driven, as is the case in our CE models. Taking into account the thermodynamic efficiencies of inner core crystallisation and latent heat release \citep[as computed in][]{Lister2003}, the dimensional, volumetric power is then $p\approx 0.2 Q_{ad}/V$. The modified, flux-based Rayleigh number can then be deduced using the proportionality exhibited in figure \ref{Rap} as $Ra^{*}_{F}=p/\gamma \rho \Omega^{3} D^{2}$, with $\gamma=0.33$. The resulting time scale ratios for Earth's core are then
\begin{eqnarray}
\left(\dfrac{\tau_{\Omega}}{\tau_{F}}\right)^{3}&=&Ra^{*}_{F}=6\te{-13}-2.5\te{-12},\label{RaEarth}\\
\dfrac{\tau_{\Omega}}{\tau_{\eta}}&=&E/Pm=10^{-9}-10^{-8},\\
\dfrac{\tau_{\eta}\tau_{\Omega}}{\tau_{A}^{2}}&=&\Lambda=5-50,\\
\dfrac{\tau_{\eta}}{\tau_{U}}&=&Rm=400-4000.
\end{eqnarray}
We see that along a sensible path connecting models and Earth, we would need to preserve at least the magnetic Reynolds number and the Elsasser number, our starting values representing reasonable midpoints in the range inferred for Earth's core. We will see below that these two conditions in fact reduce to only one once the MAC balance is considered. We thus first decide to keep $Rm$ constant. From the discussion in section \ref{MACtheory} this implies that $d_{\text{min}}$ should be close to invariant along the path. If we assume invariance for $d_{\text{min}}$, then the vorticity equivalence (\ref{vorteq}) also implies that $d_{\perp}$ is invariant. The first part of the balance (\ref{MACbal}) then reduces to $\rho\Omega U \sim g_{o} C$, and we also have $F\sim UCD^{2}$. This yields the classical thermal wind balance \citep[e.g.][]{Starchenko2002,Aurnou2003,Aubert2005,Pichon2016}, which is written here using the time scales introduced above:
\begin{equation}
\tau_{\Omega} \tau_{U}^{2} \sim \tau_{F}^{3}\label{TW}.
\end{equation}
With a constant $d_{\text{min}}$ and $f_{\text{ohm}}\approx 1$, the energy budget (\ref{Ebudget}) now \rev{becomes}
\begin{equation}
\tau_{F}^{3} \sim \tau_{A}^{2} \tau_{\eta}\label{CT04}.
\end{equation}
Note that (\ref{TW}) and (\ref{CT04}), together with  a constant $Rm=\tau_{\eta}/\tau_{U}$, indeed yield a constant Elsasser number $\Lambda = \tau_{\eta}\tau_{\Omega}/\tau_{A}^{2}$ along the path. With these relationships at hand, we may now mathematically define the rules for input parameters along the path. Any model may be characterised by a path parameter $\epsilon$ such that
\begin{equation}
Ra^{*}_{F}=\epsilon (Ra^{*}_{F})_{0}.\label{epsilondef}
\end{equation}
In the CE setup, $\epsilon$ is then an input parameter controlling the mass anomaly flux and convective power, or alternatively the rotation rate since $Ra^{*}_{F}=(\tau_\Omega/\tau_F)^{3}$. The rapidly rotating asymptotic regime is obtained for $\epsilon \rightarrow 0$. With this definition, $\epsilon=1$ corresponds to the conditions of model 0, and $\epsilon = 10^{-7}$ is appropriate to describe Earth's core conditions given the Earth values reported above. Using the definition of $Ra^{*}_{F}$ in terms of time scales, together with (\ref{TW}) and a constant $Rm=\tau_{\eta}/\tau_{U}$ we then obtain
\begin{equation}
\dfrac{E}{Pm}=\sqrt{\epsilon} \dfrac{E_{0}}{Pm_{0}}.
\end{equation}
We see here again that $\epsilon=1$ still describes model 0, while $\epsilon=10^{-7}$ indeed yields a correct Earth value $E/Pm=3.8\te{-9}$ for the magnetic Ekman number. As mentioned above, we also wish to sufficiently increase $\tau_{\nu}$ and $\tau_{\kappa}$ relative to $\tau_{\eta}$ in order to ensure a dominantly ohmic dissipation. We adopt
\begin{eqnarray}
Pm &=& \dfrac{\tau_{\eta}}{\tau_{\nu}} = \sqrt{\epsilon}~Pm_{0},\label{pathPm}\\
Pr &=& \dfrac{\tau_{\kappa}}{\tau_{\nu}}= 1,
\end{eqnarray}
such that
\begin{equation}
E= \dfrac{\tau_{\Omega}}{\tau_{\nu}} = \epsilon E_{0}.\label{pathE}
\end{equation}
We see that setting $\epsilon=10^{-7}$ implies $Pm= 7.9\te{-4}$ and $E=3\te{-12}$, meaning that although at the end of the path we have indeed rendered $\tau_{\eta}$ and $\tau_{\Omega}$ much shorter than $\tau_{\nu}$ and $\tau_{\kappa}$, we have not reached the expected Earth values for their ratios. However, we argue that $Pm$ and $E$ are small enough for thermal and viscous diffusivites to effectively become irrelevant, as is the case in Earth's core. In contrast, it is a far more important result that the path achieves a correct value of the Earth magnetic Ekman number $E/Pm$, since this number involves $\tau_{\Omega}$ and $\tau_{\eta}$, two time scales which have a major impact on the MAC balance and on energy dissipation. Other dependences on $\epsilon$ for $E$ and $Pm$ than those in equations (\ref{pathPm}, \ref{pathE}) may besides be chosen, as long as $E/Pm$ still varies like $\sqrt{\epsilon}$, and as long as $Pm$ decreases more slowly than $E^{3/4}$ along the path, an empirical condition \citep{ChristensenAubert2006} for maintaining self-sustained dynamo action that is also satisfied by (\ref{pathPm}, \ref{pathE}). \rev{For instance, perhaps a more elegant (but more computationally demanding) approach could be to simply define the path from the sole requirement to match the Earth's core values of $Ra^{*}_{F}$, $E$ and $Pm$ at $\epsilon=10^{-7}$, i.e. $Ra^{*}_{F}=\epsilon (Ra^{*}_{F})_{0}$, $E=\epsilon^{1.4}E_{0}$ and $Pm=\epsilon^{0.9}Pm_{0}$. Along this alternative path  (which satisfies all the above requirements) the constancy of $Rm$, $\Lambda$, together with the gradual enforcement of the MAC balance should then presumably emerge as results rather than prescriptions.}

With the path input parameters now fully defined, we turn to the expected scalings for the outputs along the path. We have already seen that our path implies
\begin{gather}
Rm=Rm_{0},\label{Rmscaling}\\
\Lambda=\Lambda_{0}.\label{Elscaling}
\end{gather}
The CE-type models in table \ref{table} provide a first consistency check of our theoretical approach, as $Rm$ and $\Lambda$ are indeed constant to within $\pm 10 \%$ for models chosen along the path down to $\epsilon=3\te{-4}$, and, by construction, obviously match the expected Earth values for $\epsilon=10^{-7}$. This also applies to the ST models, though with a slightly different baseline for $\Lambda$. Diffusivity-free scalings for other outputs can be derived using our path definitions (\ref{epsilondef}, \ref{pathPm}, \ref{pathE}) together with the relationships (\ref{TW}, \ref{CT04}, \ref{Rmscaling}):
\begin{eqnarray}
\dfrac{\tau_{\Omega}}{\tau_{U}}&=&Ro=\sqrt{\epsilon}~Ro_{0},\label{Roscal}  \\
\dfrac{\tau_{\Omega}}{\tau_{A}}&=&\lambda=\epsilon^{1/4} \lambda_{0}, \label{Lescal}\\
\dfrac{\tau_{A}}{\tau_{U}}&=&A=\epsilon^{1/4} A_{0}.\label{Ascal}
\end{eqnarray}
Here $Ro_{0}=1.11\te{-2}$, $\lambda_{0}=1.60\te{-2}$ and $A_{0}=0.69$ are respectively the Rossby, Lehnert and Alfv\'en numbers of model 0 (tables \ref{numbers},\ref{table}). \rev{Extrapolations of the} scalings (\ref{Roscal}-\ref{Ascal}) to Earth's core conditions ($\epsilon=10^{-7}$) \rev{yield}  $Ro=3.5\te{-6}$, $\lambda=2.8\te{-4}$ and $A=1.2\te{-2}$. These are strikingly close to the estimates $Ro=3\te{-6}$, $\lambda=2.1\te{-4}$ and $A=1.5\te{-2}$ obtained using the Earth values introduced above for $U,B,\rho,\Omega,D$. Stating the result in another way, these scalings together with the output of model 0 provide independent predictions $U=6\te{-4}\ut{m/s}$ and $B=5.3\ut{mT}$ which compare very favourably with current estimates obtained through geophysical methods \citep{Gillet2010,Buffett2010,Aubert2014}. While there is a rough equipartition of the kinetic and magnetic energy in model 0 ($A_{0}^{2}=0.48$), the predicted separation increases with decreasing $\epsilon$ and is large at Earth's core conditions, i.e. $A^{2}=\mathcal{O}(10^{-4})$.

In summary, here we have used a spatially-invariant approximation of the D13 theory to establish a smooth (i.e. devoid of abrupt physical transitions) theoretical connection between classical numerical dynamos such as model 0 and the conditions of Earth's core, along a unidimensional parameter space path constrained by the MAC balance and the need to maintain a constant value of the magnetic Reynolds number. \rev{While the extrapolations performed above are certainly supportive of this connection, the spatial invariance assumed in the path theory implies scaling exponents for the main outputs (equations \ref{Rmscaling}-\ref{Ascal}) differing from the values predicted by the (in principle) asymptotically correct D13 theory along the path. From equations (\ref{D13U}-\ref{dperpscaling},\ref{epsilondef}-\ref{pathE}), we indeed get the D13 scalings\begin{eqnarray}
Ro &=& Ro_{0}\,\epsilon^{4/9},\label{Roepsilon}\\
\lambda &=& \lambda_{0}\,\epsilon^{1/3},\label{lambdaepsilon}\\
Rm & = & Rm_{0}\,\epsilon^{-1/18},\label{Rmepsilon}\\
\Lambda & = & \Lambda_{0}\,\epsilon^{1/6},\label{Lambdaepsilon}\\
d_{\text{min}} &=&(d_{min})_{0}\,\epsilon^{1/12},\label{dminepsilon}\\
d_{\perp} &= &(d_{\perp})_{0}\,\epsilon^{1/9}.\label{mperpepsilon}
\end{eqnarray}
We note that in the case of $d_{\text{min}}$ and $d_{\perp}$, the dependences in $\epsilon$ are indeed marginal and support spatial invariance to the extent that there is less than an order of magnitude variation between $\epsilon=1$ and $\epsilon=10^{-7}$. The same also holds for $Rm$, but interestingly not for $\Lambda$. Both the D13 and spatially-invariant path theories will be checked in section \ref{res:scalings} against numerical data covering a wide portion of the path, leading to an assesment of the quality of the spatially-invariant approximation. As can be expected, we will show that the direct numerical simulation results support the D13 set of exponents (\ref{Roepsilon}-\ref{mperpepsilon}), while the results of large-eddy simulations to be introduced in section \ref{LES} are best described by the set of exponents (\ref{Rmscaling}-\ref{Ascal}), since these simulations assume spatial invariance to some extent.}

\subsection{\label{STpath}Path theory in the ST setup}
The formulation of the path theory is straightforward in the CE setup, because the convective power is in fact an input as it relates to the imposed mass anomaly flux $Ra^{*}_{F}$ (figure \ref{Rap}). The situation changes in the more widely studied ST setup, where the power is an output. Starting from model 1 (table \ref{table}, model parameters subscripted with 1 in the following), which operates with the ST setup in the same physical regime as model 0 with the CE setup, we now wish to define the path parameter $\epsilon$ in terms of the power $p$ (which is  then fully equivalent to what we did in the CE setup, equation \ref{epsilondef}):
\begin{equation}
p=\epsilon p_{1}.
\end{equation}
We wish to also keep the same dependences on $\epsilon$ for $E$, $Pm$ and $Pr$ as those introduced in (\ref{pathPm}-\ref{pathE}), i.e. $E=\epsilon E_{1}$, $Pm=\sqrt{\epsilon}~Pm_{1}$, and $Pr=1$. To obtain the path rule for the input parameter $Ra^{*}$, we recall the proportionality between the convective power and the mass anomaly flux (figure \ref{Rap}), which in the ST setup \rev{is} \citep{ChristensenAubert2006}:
\begin{equation}
p \sim Ra^{*} (Nu-1) \dfrac{E}{Pr}.
\end{equation}
We thus need a relationship between the Nusselt and Rayleigh numbers. \rev{Such a} relationship usually involves the canonical Rayleigh number $Ra$ and its critical value for convection onset $Ra_{c}\sim E^{-4/3}$ \citep{Busse1970}. It can generically be expressed as $Nu-1 \sim (Ra/Ra_{c})^{\beta} \sim Ra^{\beta} E^{4\beta/3}$, and with equation (\ref{cantomod}) and the fact that $Pr=1$ this yields $Nu-1 \sim (Ra^{*})^{\beta} E^{-2\beta/3}$. The convective power then scales with $p \sim (Ra^{*})^{\beta+1} E^{-2\beta/3+1}$, and the requirements $p \sim \epsilon$ and $E \sim \epsilon$ thus imply $Ra^{*}=\epsilon^{2\beta/3(\beta+1)}~Ra^{*}_{1}$. Rotating convection and dynamo studies suggest values ranging from $\beta=6/5$ \citep{ChristensenAubert2006,Aurnou2007,King2010} to $\beta=3/2$ \citep[the diffusivity-free prediction,][]{Gillet2006,Julien2012,Stellmach2014}. These yield very similar dependences $\epsilon^{4/11}$ to $\epsilon^{2/5}$ for $Ra^{*}$. In the following we choose to retain the simpler expression 
\begin{equation}
Ra^{*}=\epsilon^{2/5} Ra^{*}_{1},
\end{equation}
for completing the definition of the ST path. We note that the canonical Rayleigh number $Ra$ then increases with $\epsilon^{-8/5}$, which is steeper than the increase of the critical value $Ra_{c}\sim \epsilon^{-4/3}$, meaning that supercriticality indeed increases as we progress along the path. 

Direct numerical simulations (table \ref{table}, ST/D cases) show that the obtained $\epsilon$, reported in table \ref{table} as the ratio $p/p_{1}$, very closely matches the intended $\epsilon$ defined as $E/E_{1}$. This provides a posterior check for our analysis. For the large-eddy simulations (ST/L cases) to be introduced in section \ref{LES}, however, the \rev{obtained value is lower than the intended value, and the difference increases} with decreasing $\epsilon$. The hyperdiffusive treatment of the velocity and density anomaly fields \rev{indeed leads to a loss of the convective power carried through} the interaction of small-scale velocity and density anomaly. \rev{Note though that this problem is specific to the ST setup, since in the CE setup the relationships $p/p_{0}\approx Ra^{*}_{F}/(Ra^{*}_{F})_{0}=E/E_{0}=\epsilon$ hold with markedly better accuracy (table \ref{table}).}

\subsection{\label{LES}Large-eddy simulations}
Exhibiting numerical solutions for low values of $\epsilon$ is difficult because hydrodynamic turbulence sets up as one progresses along the path, as witnessed by the Reynolds number scaling that can be derived from (\ref{Roscal}):
\begin{equation}
Re=Re_{0} \epsilon^{-1/2},\label{Rescal}
\end{equation} 
with $Re_{0}=372$ in the CE case. As a result, the exploration of the path with a fully resolved direct numerical simulation (DNS) is currently possible only down to $\epsilon=0.1$ (cases CE/D and ST/D in table \ref{table}). Achieving lower values of $\epsilon$ requires to define relevant large-eddy simulations (LES, cases CE/L and ST/L in table \ref{table}). We have seen in section \ref{MACtheory} that \rev{the essential part of the energy transfers should occur between the scales $d_{\perp}$ and $d_{\text{min}}$, with $d_{\perp}>d_{\text{min}}$ and the separation between the two being mainly controlled by the level of magnetic turbulence $Rm$ (equations \ref{dminscaling},\ref{dperpscaling}). We have also} seen that we could approximate $d_{\text{min}}$ by a constant along a path of constant $Rm$, implying that $d_{\perp}$ is also a constant. \rev{These points suggest} that the hydrodynamic turbulence gradually setting up at scales below $d_{\text{min}}$ as we progress along this path is irrelevant to the determination of the large-scale structure. This latter structure is indeed controlled by a large-scale MAC balance at scale $d_{\perp}$ and only a small part of the convective power does cascade further below the scale $d_{\text{min}}$, the major part being dissipated there through ohmic losses.

We may thus parameterise the hydrodynamic turbulence below scale $d_{\text{min}}$ without too much loss in physical relevance and accuracy, and our numerical computation may be restricted to the sole length scale range $d_{\text{min}}\le d \le  D$, thus alleviating the common scale-separation problems encountered at extreme control parameter values. Using $d_{\text{min}} \approx \sqrt{1/\,2Rm}$ \citep[][a relationship also well verified in our data set, see table \ref{table}]{Christensen2004} and $Rm=1000$, this corresponds to a numerical expansion up to spherical harmonic degree $l_{\text{max}}=140$. In CE/L and ST/L large-eddy cases (table \ref{table}) we have used either $l_{\text{max}}=133$ or $l_{\text{max}}=170$ for reasons of numerical efficiency. We adopt a hyperdiffusive treatment recently proposed by \cite{Nataf2015} that we apply to the diffusion of momentum and density anomaly, but not to that of the magnetic field, which remains fully resolved. In the numerical implementation, the principle is to use effective diffusivities $\nu_\text{eff},\kappa_\text{eff}$ that depend on the harmonic degree $l$ and the molecular diffusivities $\nu,\kappa$ according to
\begin{eqnarray}
(\nu_\text{eff},\kappa_\text{eff})&=&(\nu,\kappa)~\mathrm{for}~l < l_{h}, \\
(\nu_\text{eff},\kappa_\text{eff})&=&(\nu,\kappa)\, q_{h}^{l-l_{h}} ~\mathrm{for}~l\ge l_{h}.
\end{eqnarray}
Here $l_{h}$ is the cut-off degree below which the hyperdiffusive treatment is not applied. For this treatment to remain physically accurate, it should satisfy the following requirements: {\it (i)} it should not perturb the large-scale MAC balance, i.e. $l_{h}$ should be sufficiently larger than $l_{\perp}=\pi D/d_{\perp}$, and {\it (ii)} the additional viscous energy losses that it implies should remain small relatively to ohmic losses. These requirements are satisfied here by prescribing $l_{h}=30$, a value which, as we will see in section \ref{res:forcebal}, is three times larger than our estimate $l_{\perp}\approx 10$, and values of $q_{h}$ close to 1 (table \ref{table}), enabling a very smooth increase of hyperdiffusivity with harmonic degree. In section \ref{results} we will validate this treatment by comparing DNS and LES simulations in the range where both are feasible, and by demonstrating that the additional viscous losses are indeed negligible in the asymptotic limit. There is some leeway in the choices of $q_{h}$ and $l_{h}$. For the CE/L case performed at $\epsilon=0.1$ we have indeed checked that varying $q_{h}$ in the range $1.02\le q_{h} \le 1.07$ yields similar results to within 5\%. The same holds when varying $l_{h}$ in the range $15\le l_{h} \le 60$ in the CE/L case performed at $\epsilon=0.33$. Given the numerical schemes used to solve the equations, the hyperdiffusive treatment only applies to the lateral directions of the numerical calculation, making it still necessary to increase the radial resolution as $\epsilon$ decreases. This is however a far more tractable numerical problem than that of expanding the grid in all three spatial directions. Combined with the fact that we avoid treating viscous boundary layers, this permitted to calculate CE cases down to $\epsilon=3.33\te{-4}$ and an Ekman number $E=10^{-8}$. This is the lowest Ekman number reached to date in self-sustained spherical convective dynamos, though it should be acknowledged that this value applies to the degree range $l\le30$ only, after which the effective Ekman number increases. 

\section{\label{results}Numerical results}
\subsection{\label{res:forcebal}Spatial structure of the asymptotic MAC force balance}

\begin{figure}
\centerline{\includegraphics[width=13.5cm]{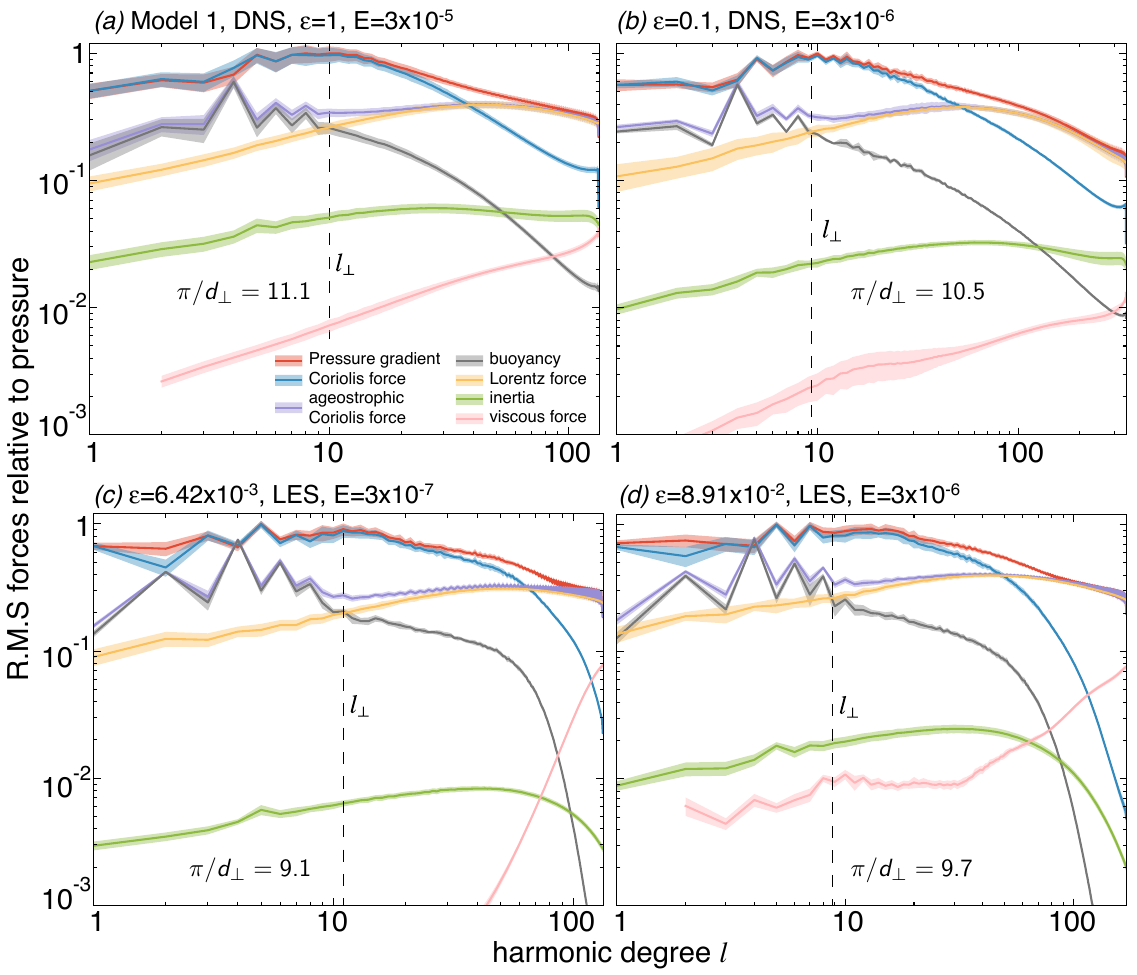}}
\caption{\label{forcebal} Force balance spectrum in the ST numerical simulations (viscous boundary layers excluded), presented as a function of the spherical harmonic degree $l$, and normalised relative to the \rev{peak} amplitude of the pressure gradient, for two DNS ({\it a,b}) and two LES ({\it c,d}) cases performed at varying $\epsilon$ along the path. Panels ({\it b}) and ({\it d}) present a comparison of DNS and LES at a constant $\epsilon$ value. Note that the maximal spherical harmonic degree $l_{\text{max}}$ of the calculations differs in the four cases (see table \ref{table}). The solid lines represent the time-averaged force balance, and the shaded regions represent the $\pm 1$ std. dev. in the instantaneous force balance. The harmonic degree $l_{\perp}$ is defined as the intersection between the buoyancy and Lorentz forces, such that these two forces equally contribute to balance the part of the Coriolis force not balanced by the pressure gradient. \rev{For comparison purposes, the predictions $\pi/d_{\perp}$ obtained from equation (\ref{dperp}) using values in table \ref{table} are also reported in panels ({\it a-d}).}}
\end{figure}

In figure \ref{forcebal} we first analyse the force balance operating in the ST simulations (similar results hold for the CE simulations). To this end, for each force $\vc{f}$ we follow \cite{Soderlund2012,Soderlund2015} and compute
\begin{equation}
f_{rms}^{2}=\dfrac{1}{V}\int_{V} \vc{f}^{2}\,\mathrm{d}V=\sum_{l=0}^{l_{\text{max}}} f_{l}^{2}.
\end{equation}

\rev{There are a few} novelties of our calculation relative to \cite{Soderlund2012,Soderlund2015}. \rev{First}, we have access to the part of the Coriolis force not balanced by the pressure gradient \rev{(the ageostrophic part)}. \rev{Second, we reveal the actual force balance in the bulk of the fluid by} excluding viscous boundary layers from the calculation. \revv{Note that this methodology is similar to that used in \cite{Yadav2016PNAS} and the results are hence directly comparable. A third point which is original to the present work is that} we decompose $f_{rms}^{2}$ into a sum of contributions $f_{l}^{2}$ along spherical harmonic degrees and present $f_{l}$ as a function of $l$. \rev{At the order following that of the diagnostic balance between Coriolis and pressure forces,} a MAC balance is robustly observed in all simulations between \rev{the ageostrophic} Coriolis, buoyancy and Lorentz forces. This force balance is \rev{structurally} similar among all simulations (compare the four panels of figure \ref{forcebal}). The harmonic degree $l_{\perp}\approx 10$ corresponding to \rev{an optimal MAC balance} may be identified as the intersection between the contributions of the Lorentz and buoyancy forces, and is largely invariant, both in the DNS and LES simulations. For harmonic degrees smaller than $l_{\perp}$ (length scales larger than $d_{\perp}$) the main balance is between the buoyancy and \rev{ageostrophic} Coriolis forces, and the Lorentz force substitutes to the buoyancy force for harmonic degrees larger than $l_{\perp}$ (length scales smaller than $d_{\perp}$). The temporal variability of this basic structure is quite low, implying that the MAC balance holds at all instants in the numerical simulations. \rev{Independent predictions $l_{\perp}=\pi/d_{\perp}$ (reported in figure \ref{forcebal}) can be obtained from equation (\ref{vorteq}) and the values in table \ref{table}, where the dimensionless D13 length scale $d_{\perp}$ is
\begin{equation}
d_{\perp} = f_{\text{ohm}}d_{\text{min}}^{2} Rm.\label{dperp}
\end{equation}
Note that (\ref{dperp}) reintroduces the contribution of $f_{\text{ohm}}$ that should arise from (\ref{Ebudget}) but has been neglected in (\ref{vorteq}). These predictions are essentially invariant and closely match the measure obtained by the crossings of forces in figure \ref{forcebal}. This confirms that $l_{\perp}$ indeed corresponds to the D13 scale $d_{\perp}$. This also demonstrates the conjecture that the large scale at which the convective dynamo is organised is indeed controlled by the vorticity equivalence (\ref{vorteq}) that follows from the MAC balance.} It is interesting to note that the starting point of the ST path (model 1, figure \ref{forcebal}{\it a}) already exhibits a well-respected MAC balance, though the contributions of inertial and viscous forces are still sizeable. Decreasing $\epsilon$ by a factor 10 in a DNS simulation (figure \ref{forcebal}{\it b}) decreases the contribution of inertia by a factor 3 roughly, as expected from the predicted Rossby number scaling (equation \ref{Roscal}) along the path. The gap between inertial and viscous forces also enlarges, though at a pace somewhat slower than that predicted by Reynolds number scaling (\ref{Rescal}), hinting at a non-trivial length scale for viscous dissipation which we leave outside the scope of the present study.

The agreement between DNS and LES performed at a similar value of $\epsilon$ (figure \ref{forcebal}{\it b,d}) is excellent regarding the harmonic structure and relative amplitudes of pressure, Coriolis, buoyancy, Lorentz and inertial forces up to degree $l\le 70$. The LES obviously shows an increase in viscous forces after degree $l_{h}=30$ but it is worth noting that up to $l=l_{\text{max}}$ these do not reach a first-order dominance, such that the nature of the small-scale force balance is also preserved. An interesting point is that for degrees $l\le l_{h}$, the viscous force also slightly increases in the LES relatively to the DNS. Finally, comparing the DNS case with $\epsilon=1$ to the LES case with $\epsilon=6.42\te{-3}$ (figure \ref{forcebal}{\it a,c}), we see that despite the use of hyperdiffusivity, this latter simulation has reached a nearly inviscid regime in the harmonic degree range $l\le 50$, with the viscous force at least two orders of magnitude lower than the MAC forces. In this harmonic degree range, inertia is also rendered largely subdominant, with its typical amplitude one to two orders of magnitude below the MAC forces. All these points underline the success of the path and LES approaches in reaching the MAC balance. This regime has been observed down to the lowest value of $\epsilon$ that we could achieve, halfway in logarithmic distance between the classical models and the Earth conditions along the parameter path. These results provide strong support to the claims that the MAC balance is indeed the planetary, asymptotic regime, and that our numerical models are close to this regime. 

\begin{figure}
\centerline{\includegraphics[width=13.5cm]{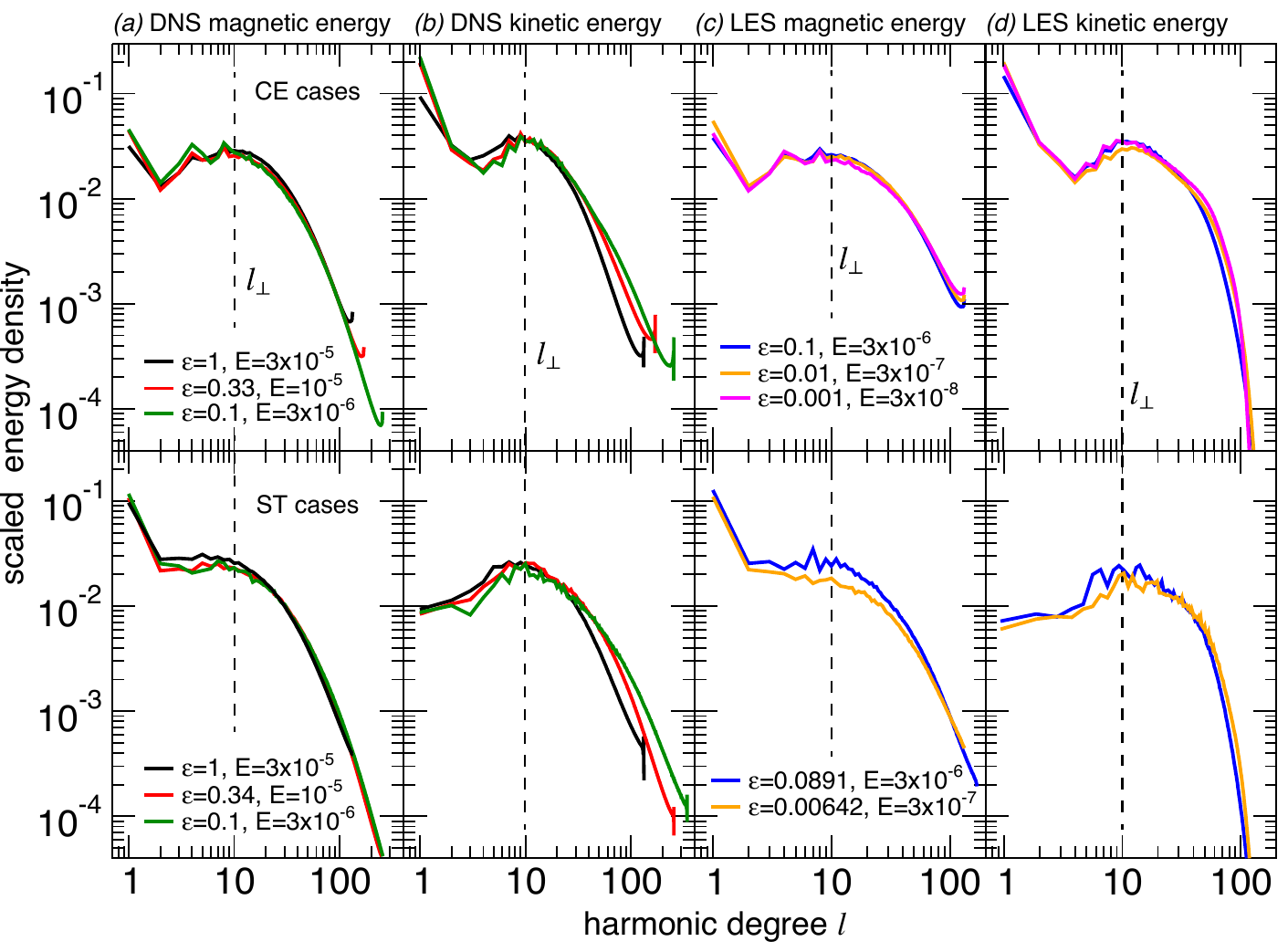}}
\caption{\label{spec}Time-averaged magnetic ({\it a,c}) and kinetic ({\it b,d}) energy spectra in the CE (top row) and ST (bottom row) setups, for DNS ({\it a,b}) and LES ({\it c,d}) simulations performed at varying $\epsilon$. To obtain the normalised magnetic energy spectra in panels ({\it a,c}), the magnetic field amplitude is normalised by its scaling prediction $\lambda_{0,1} \epsilon^{1/4}$ (equation \ref{Lescal}), where $\lambda_{0,1}$ stands for the Lehnert number of model 0 (CE) or 1 (ST). Similarly, in panels ({\it b,d}) the amplitude of the velocity field is also normalised by $Ro_{0,1} \epsilon^{1/2}$ (equation \ref{Roscal}). The total normalised energy is close to 1 in all cases. For reference, the dashed vertical lines mark the harmonic degree $l_{\perp}$ corresponding to $d_{\perp}$ as identified in figure \ref{forcebal}.}
\centerline{\includegraphics[height=5.5cm]{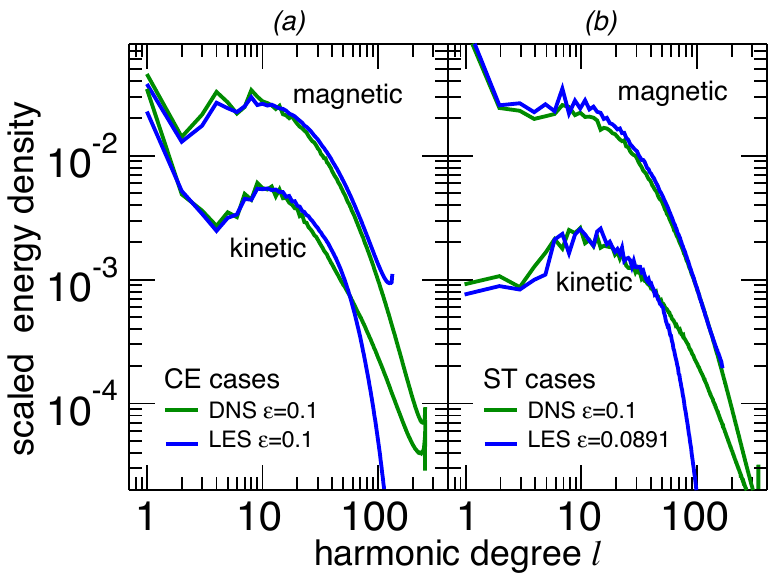}}
\caption{\label{specdnsvsles}Comparison of time-averaged energy spectra obtained in DNS and LES at a given $\epsilon$ value corresponding to an Ekman number $E=3\te{-6}$, for CE ({\it a}) and ST ({\it b}) cases. The amplitude of magnetic and velocity fields are both normalised by $\lambda_{0,1} \epsilon^{1/4}$. The total normalised magnetic energy is then close to 1 while the total normalised kinetic energy is close to the separation factor $A^{2}=0.2$ (CE) and $A^{2}=0.14$ (ST).}
\end{figure}

\subsection{\label{res:structure}Spatial invariance of the solutions along the path}

We now turn to the analysis of the spatial structure of the fields along the path. The large-scale invariance can be first demonstrated by examining energy spectra from direct numerical simulations (figure \ref{spec}{\it a,b}) performed in the range $0.1\le\epsilon\le1$. The magnetic spectral energy density profile (figure \ref{spec}{\it a}) is \rev{structurally invariant} in the range $l \le l_{diss}$, where $l_{diss} \approx 133$ is the harmonic degree corresponding to the magnetic dissipation length scale $d_{\text{min}}$. \rev{The spectral profile of kinetic energy density (figure \ref{spec}{\it b}) is also invariant in the large-scale range $l\le 30$ but shows an enrichement in smaller scales as $\epsilon$ decreases.} Large-eddy simulations (figure \ref{spec}{\it c,d})  present even higher levels of \rev{structural} similarity, both for the velocity and magnetic fields, throughout the resolved spectral range. CE-type models are generally more \rev{invariant} than ST-type models, owing to their better control on $\epsilon$. Comparing a DNS and a LES performed at a similar value of $\epsilon$ (figure \ref{specdnsvsles}) it is clear that despite using a significantly lower value of $l_{\text{max}}$, the LES accurately captures the \rev{spectral distribution of magnetic energy} almost up to degree $l_{\text{max}}=133$, and the \rev{kinetic energy distribution} at least up to degree 30. These results are found to hold irrespectively of the boundary condition choice, with CE and ST yielding comparable energy distributions, at the exception of velocity for $l=1$ which is enhanced in the CE setup by an inhomogenenous mass anomaly forcing at the inner boundary \citep{Aubert2013b}. In all cases, the kinetic energy spectra feature a peak around the value $l_{\perp}\approx 10$ already identified in the force balance (figure \ref{forcebal})\rev{, implying a predominant organisation of convective structures around this scale.}

\begin{figure}
\centerline{\includegraphics[height=13cm]{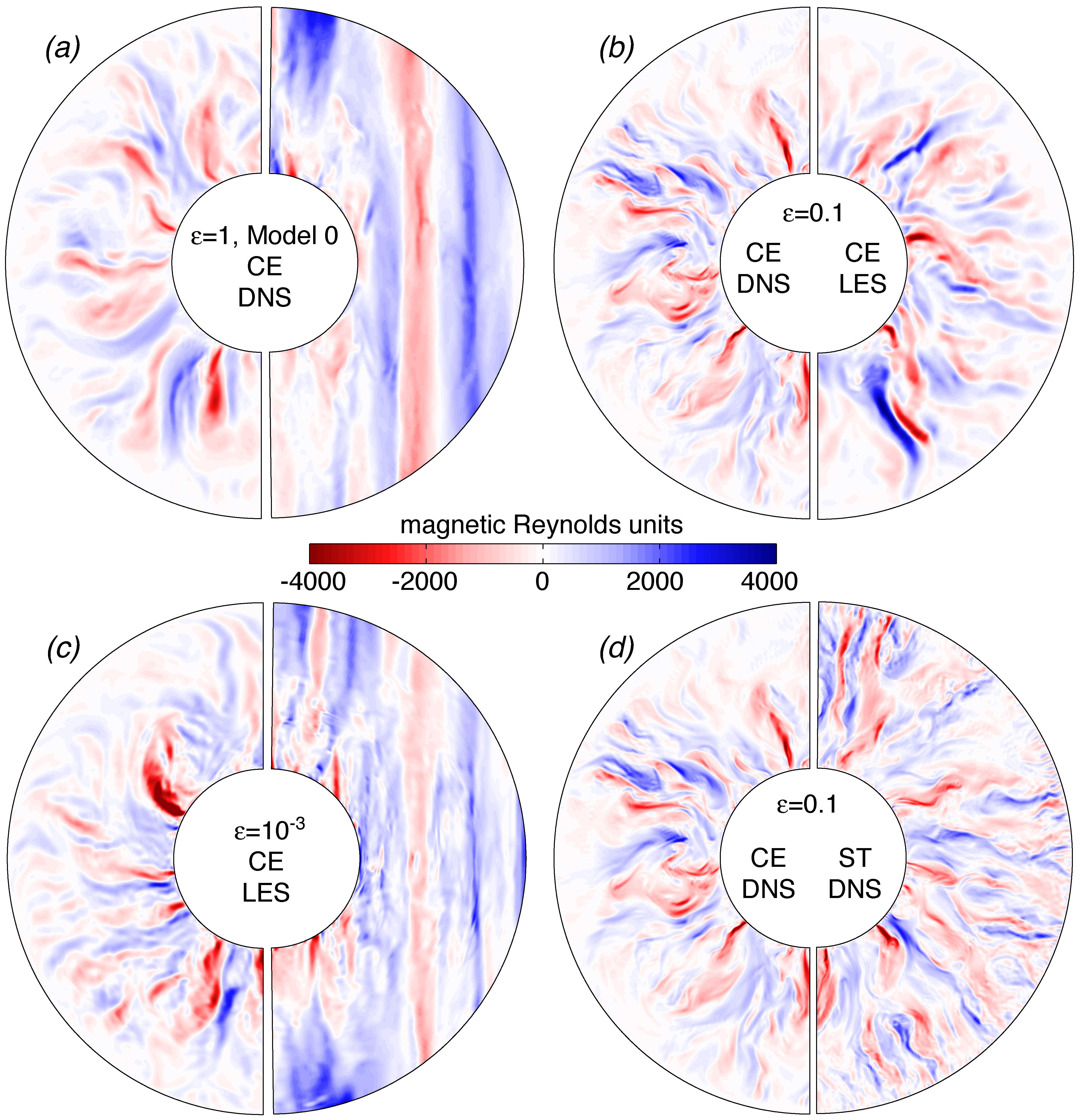}}
\caption{\label{Vplanforms} Evolution of velocity planforms (snapshots, units of $\eta/D$ or magnetic Reynolds number units) in CE-type calculation as $\epsilon$ decreases ({\it a-c}), and comparisons at fixed $\epsilon$ between DNS and LES ({\it b}), and between CE and ST cases ({\it d}). In panels ({\it a}) and ({\it c}), the left and right panels respectively present snapshots of the radial velocity in a half equatorial plane (red is outwards), and a meridional cut of the azimuthal velocity (blue is westwards). In ({\it b}) and ({\it d}) all panels represent snapshots of the equatorial radial velocity.}
\end{figure}

\begin{figure}
\centerline{\includegraphics[width=13.5cm]{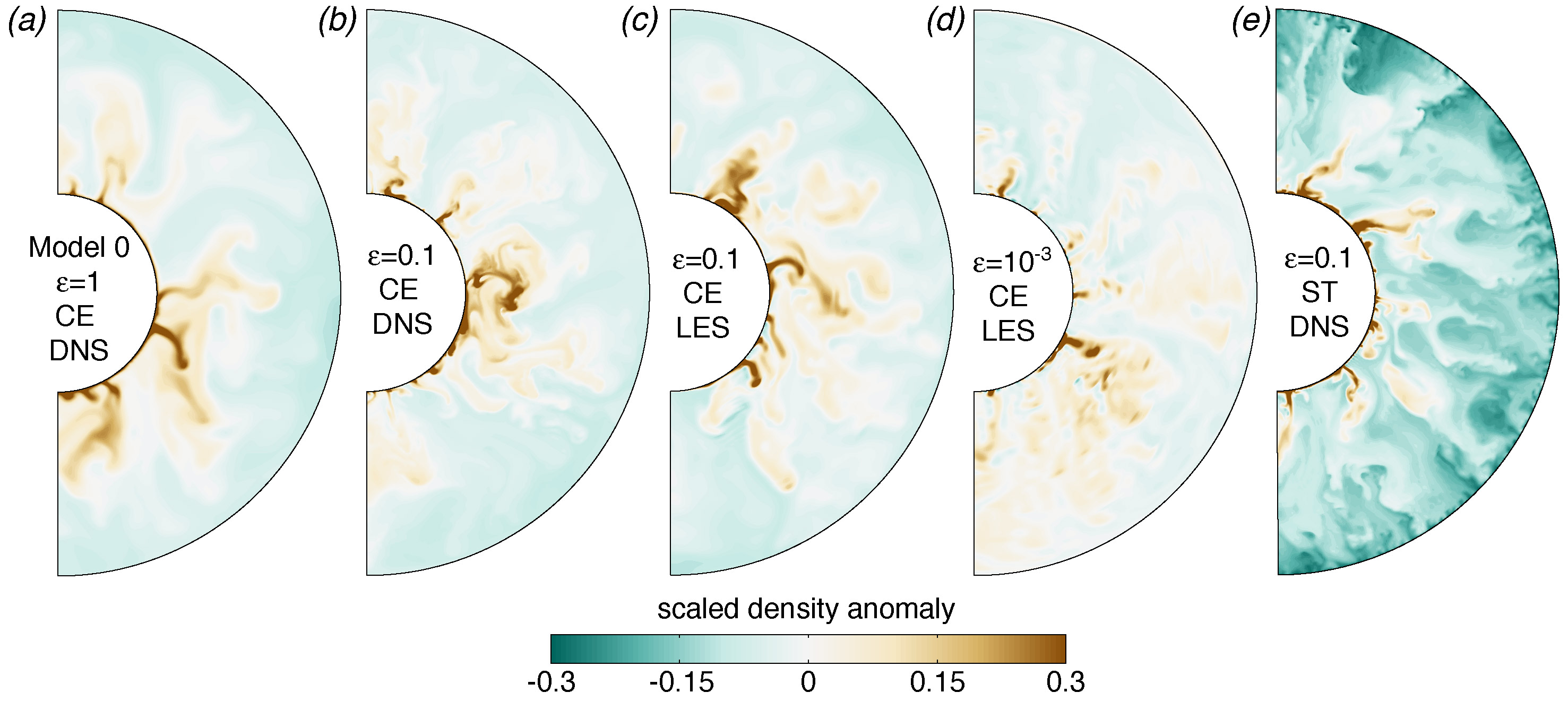}}
\caption{\label{Tplanforms}Evolution of equatorial density anomaly planforms (half-plane snapshots, orange denotes lighter fluid) as $\epsilon$ decreases ({\it a-d}) and comparisons at fixed $\epsilon$ between DNS and LES ({\it b,c}) and between CE and ST cases ({\it b,e}). In CE cases ({\it a-d}) the dimensionless density anomaly $C$ is normalised by $C_{0} \epsilon^{-1/2}$, with $C_{0}=6000$. In the ST case ({\it e}) $C$, is normalised by $Pr/E$ with $Pr=1$ and $E=3\te{-6}$.}
\end{figure}

\rev{To check how these spectral results translate into the physical space, we further examine equatorial and meridional velocity planforms in the CE setup (figure \ref{Vplanforms}{\it a-c}). From the kinetic energy spectra observed in figure \ref{spec}, we expect the equatorial planforms of radial velocity to remain structurally similar as $\epsilon$ varies, particularly in the large-scale range where an azimuthal wavenumber $m\approx l_{\perp}\approx 10$ should emerge. Comparing the CE-type DNS simulations at $\epsilon=1$ and $\epsilon=0.1$ (figure \ref{Vplanforms}{\it a,b}, left panels), we indeed observe the preservation of structures at this large scale, while a decrease in $\epsilon$ leads to the refinement of small-scale details. Compared to its LES counterpart, the DNS at $\epsilon=0.1$ (figure \ref{Vplanforms}{\it b}) also refines the small-scale details while not changing the large-scale structure, thus visually confirming the relevance of the LES. Comparing equatorial planforms obtained in the CE and ST DNS setups (figure \ref{Vplanforms}{\it d}), we find structural similarity close to the inner boundary and a richer small-scale content in the ST setup close to the outer boundary \citep[as previously shown by][]{Sakuraba2009}. This reflects a fundamental change in how convective power is transported. In the ST setup, the part of the convective power carried by these small-scale velocity structures together with their corresponding density anomalies (see figure \ref{Tplanforms}{\it e}) is not accounted for in the LES (compare the values of $p$ in table \ref{table}), leading to deviations from the ideal path (see section \ref{STpath}) and to a decrease of the LES accuracy. In contrast, CE-type LES simulations preserve the DNS convective power remarkably well (compare again $p$ values in table \ref{table}), save for a slight change of mixing properties (see trend for $\gamma=p/Ra^{*}_{F}$ in figure \ref{Rap}{\it b}) that should remain negligible at Earth's core conditions. This confirms that convective power is mostly transferred at large scales in the CE setup.}

\rev{Turning now to meridional planforms,} the axially columnar structure of the azimuthal velocity field (figure \ref{Vplanforms}{\it a,c}, right panels) outside the tangent cylinder is also preserved as $\epsilon$ decreases, together with retrograde (westward) polar vortices at the upper and lower ends of the tangent cylinder, confirming the ansatz $d_{\parallelsum}\sim D$ made in section \ref{path}. The power input in the system indeed increases as $\epsilon$ decreases, but so does also the rotational constraint, leading to a preservation of the columnar structure. Types CE and ST have zonal flows of similar amplitude despite the change in boundary conditions from stress-free to rigid \citep[not shown, previously already documented by][]{Yadav2013FS}. \cite{Livermore2016} have suggested that the amplitude of zonal flows may asymptotically scale differently depending on mechanical boundary conditions if these are limited by the residual viscosity. This does not apply here because zonal flows in spherical, convective, dipole-dominated dynamos such as those discussed here are thermal-wind limited \citep{Aubert2005}. 

\rev{Equatorial planforms of the density anomaly} (figure \ref{Tplanforms}) also lead to similar observations, \rev{with broadly structurally similar patterns at large scales} (figure \ref{Tplanforms}{\it a-d}) and \rev{the DNS simply refining the picture obtained with the LES} (figure \ref{Tplanforms}{\it b,c}). Both CE and ST setups produce dominant bottom-originated convective plumes (figure \ref{Tplanforms}{\it b,e}), as theoretically expected, but the ST setup additionally produces small-scale, secondary return plumes originated at the outer boundary. In the CE setup, the dimensionless density anomaly $C$ is found to scale with $C_{0}\epsilon^{-1/2}$, as can be predicted from the expression of convective power (\ref{pdef}), together with the velocity scaling (\ref{Roscal}) and the fact that the ratio $p/Ra^{*}_{F}$ is constant (figure \ref{Rap}), and we find $C_{0}=6000$. The typical dimensional density anomaly $\delta C$ in Earth's core may then be estimated through the following diffusivity-free, power-driven scaling obtained from equations (\ref{fluxra}) and (\ref{epsilondef}):
\begin{equation}
\delta C/\rho = C_{0} (Ra^{*}_{F})_{0} \dfrac{\Omega^{2} D}{g_{0}}\epsilon^{1/2},\label{Cscaling}
\end{equation}
with $C_{0} (Ra^{*}_{F})_{0}=0.162$. Using $\epsilon=10^{-7}$ together with the previously introduced values $D=2260 \ut{km}$, $\Omega=7.29\te{-5} \ut{s^{-1}}$, and $g_{0}=10 \ut{m/s^{2}}$, this yields $\delta C/\rho=6\te{-8}$. Converting this anomaly into a temperature deviation $\delta T$ from the core adiabat, we obtain $\delta T=\delta C / \alpha\rho=6\te{-3} \ut{K}$, where $\alpha=10^{-5} \ut{K^{-1}}$ is the thermal expansion coefficient. Either $\delta C$ or $\delta T$ are in good agreement with geophysical estimates \citep[e.g.][]{Aurnou2003,ChristensenAubert2006}, which again underlines the relevance of the path chosen to connect model and Earth's core conditions. 

\begin{figure}
\centerline{\includegraphics[width=13cm]{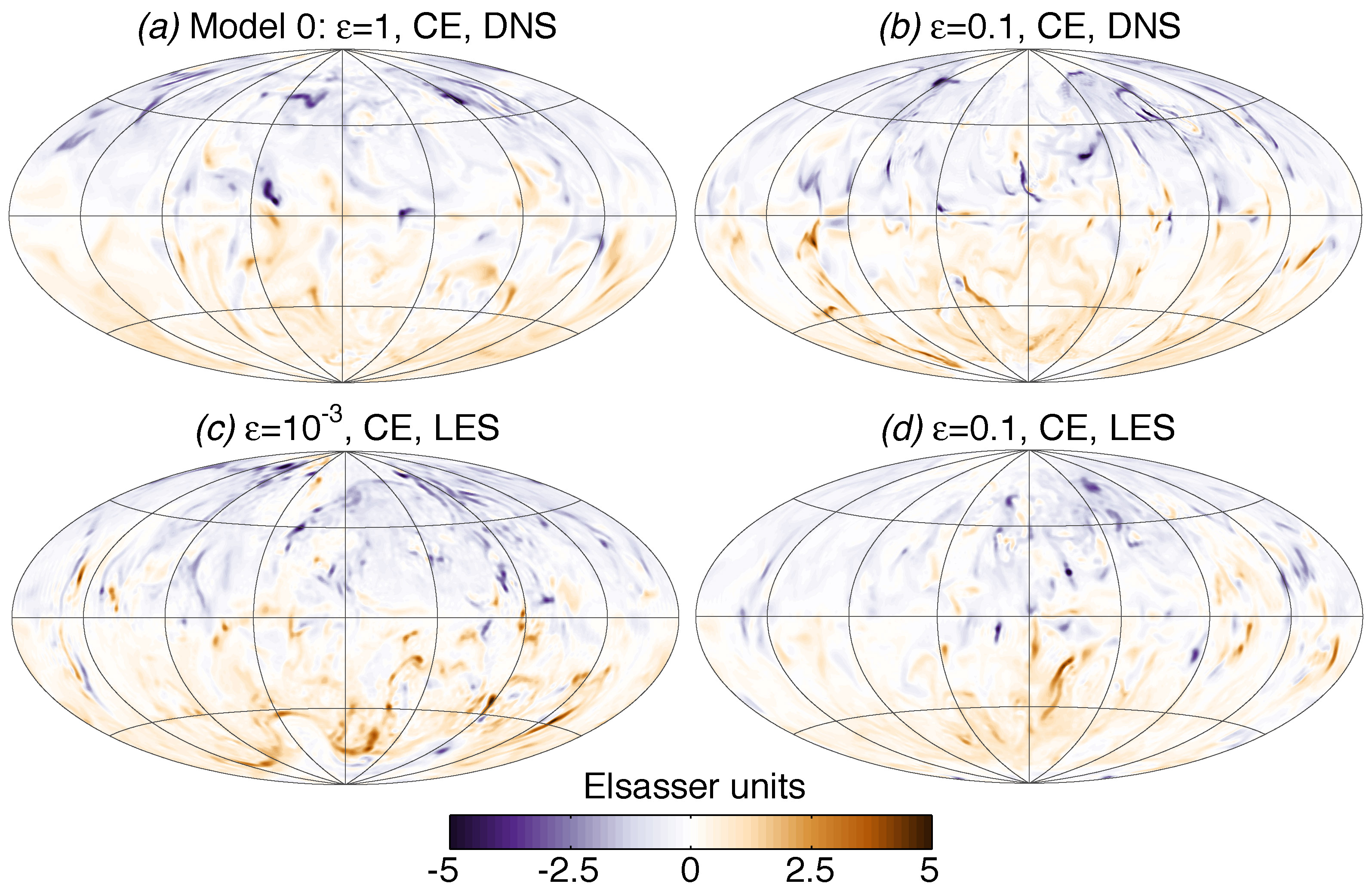}}
\caption{\label{Bplanforms}Evolution of the radial magnetic field pattern at the model outer boundary (snapshots, units of $\sqrt{\rho\mu\eta\Omega}$ or Elsasser units, orange is outwards) in the CE setup as $\epsilon$ decreases ({\it a-c}), and comparison at fixed $\epsilon$ between DNS and LES ({\it b,d}).}
\centerline{\includegraphics[width=13cm]{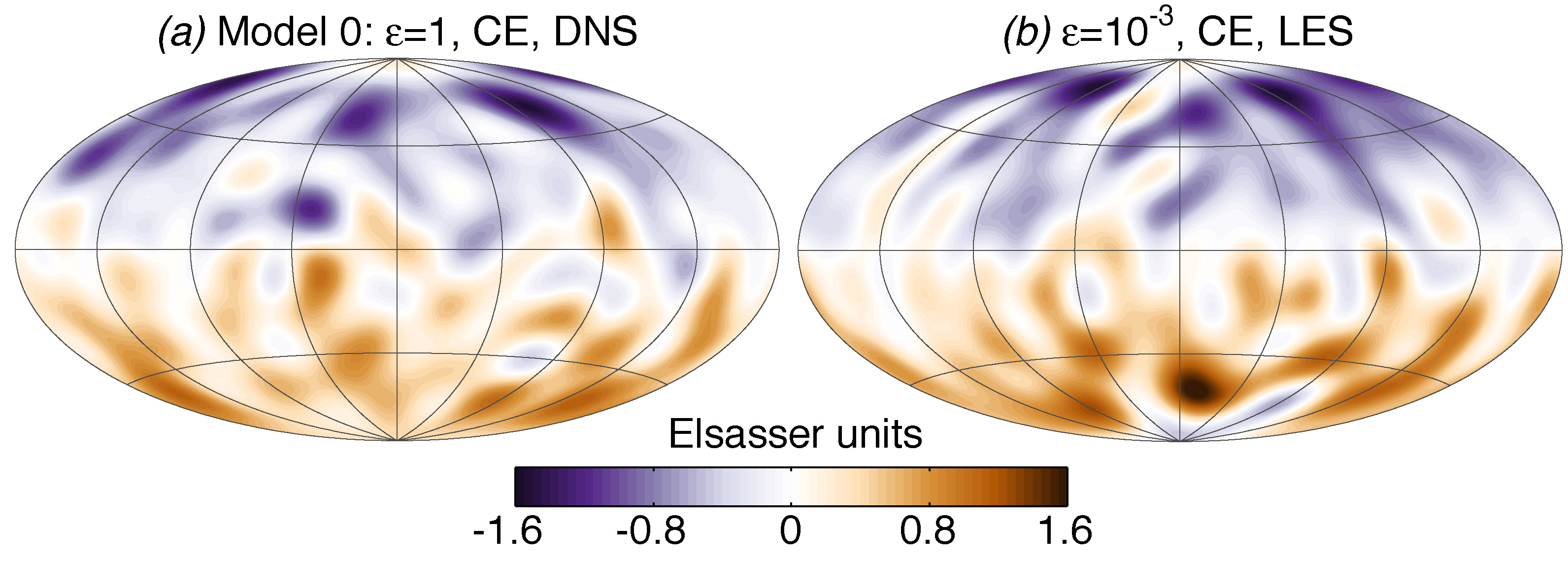}}
\caption{\label{Bplanforms13}Radial magnetic field pattern at the model outer boundary (snapshots, units of $\sqrt{\rho\mu\eta\Omega}$ or Elsasser units, orange is outwards), filtered below spherical harmonic degree and order 13, in the CE setup for a DNS at $\epsilon=1$ ({\it a}), and a LES at $\epsilon=10^{-3}$ ({\it b}).}
\end{figure}

Figures \ref{Bplanforms},\ref{Bplanforms13} present maps of the radial magnetic field at the model outer boundary for CE-type calculations. The full-resolution magnetic field patterns (figure \ref{Bplanforms}) are again remarkably \rev{invariant}, with an axial dipole-dominated morphology, high-latitude flux patches of normal polarity at the intersection of the outer boundary with the tangent cylinder, westward-drifting, low-latitude patches of normal polarity, a magnetic flux deficit inside the tangent cylinder due to the action of polar vortices, and localised inverse flux patches. The good results obtained in the original CE dynamo model \citep[similar to model 0, figure \ref{Bplanforms}{\it a}, see also][]{Aubert2013b} as regards the static and kinematic similarity of the model output to the geomagnetic field are thus preserved, and demonstrated here to pertain to the asymptotic regime. Figure \ref{Bplanforms13} further shows that when filtered to the typical resolution $l\le 13$ of modern geomagnetic observations, the model output preserves the Earth-like morphological properties \citep[as objectively defined in][]{Christensen2010} of the initial model 0. We also note in figures \ref{Bplanforms},\ref{Bplanforms13} that the field amplitude presented in Elsasser units remains constant as $\epsilon$ decreases, in agreement with equation (\ref{Elscaling}). Finally, considering the ratio of the root-mean-squared magnetic field inside the shell to that at the outer boundary, the CE models yield a constant value of about 5 if the full outer boundary magnetic field is considered, or 7 if that field is truncated to spherical harmonic degree 13. This latter value is in agreement with geophysical estimates \cite[e.g][]{ChristensenAubert2006,Gillet2010,Buffett2010}. 

\begin{figure}
\centerline{\includegraphics[width=13cm]{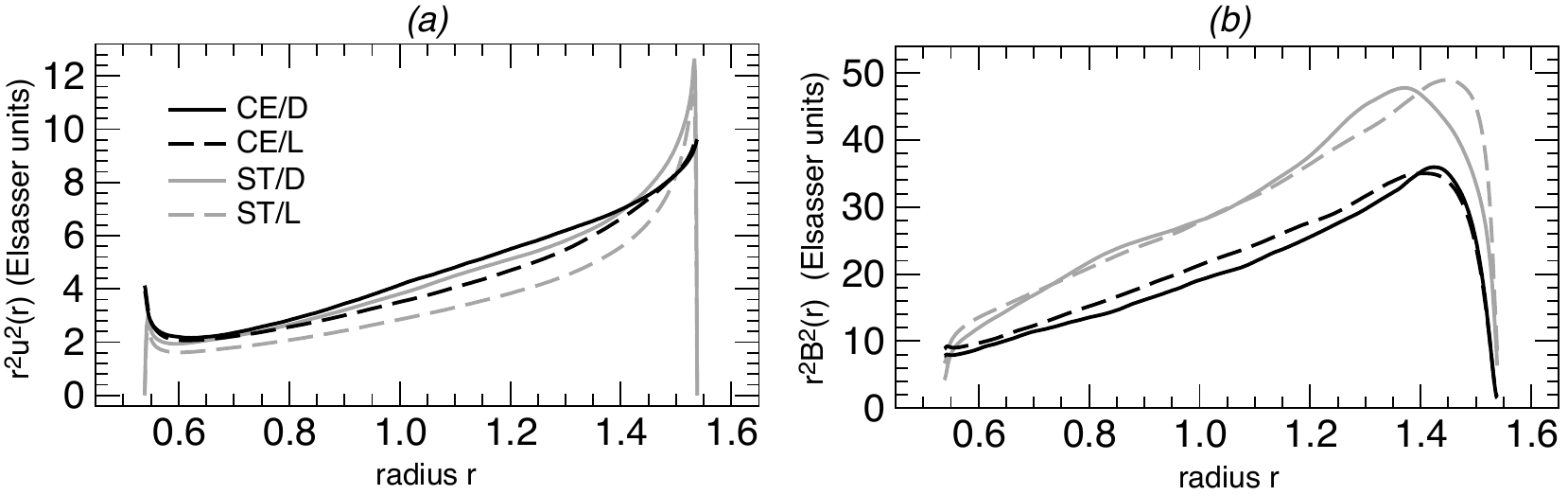}}
\caption{\label{Eprofiles} Spherical- and time-averaged radial \rev{distributions of the kinetic and magnetic energies} $r^{2}u^{2}$ ({\it a}) and $r^{2}B^{2}$ ({\it b}), for CE models (black) computed at $\epsilon=0.1$, and ST models (grey) computed at $\epsilon=0.1$ (DNS) and $\epsilon=0.0891$ (LES). DNS (solid lines) and LES (dashed lines) calculations are presented. Velocity and magnetic fields are presented in units of $\sqrt{\eta \Omega}$ and $\sqrt{\rho \mu \eta \Omega}$ (Elsasser units), respectively, such that the radial integral of the magnetic energy density is the Elsasser number $\Lambda$ and the radial integral of the kinetic energy density is $A^{2}\Lambda$, revealing the energy separation factor $A^{2}$.}
\end{figure}

We finish our structural analysis by examining radial profiles of the kinetic and magnetic \rev{energies} (figure \ref{Eprofiles}) at a fixed $\epsilon$ value, for DNS and LES simulations carried out in the CE and ST setups. These profiles show that the choice of boundary conditions has a weak influence on the distribution of energy in the shell. Indeed, the ST and CE kinetic energy profiles (figure \ref{Eprofiles}{\it a}) are strikingly similar, with the only noticeable differences residing in the ST viscous boundary layers that are absent in the CE setup. Likewise, the ST and CE magnetic energy profiles (figure \ref{Eprofiles}{\it b}) only differ in their amplitude, as already noted in table \ref{table} (see also figure \ref{Bscalings}{\it b}). Turning now to the agreement between DNS and LES, the best results are obtained in the CE setup, while in the ST setup the discrepancies are most visible near the outer boundary. There, the inhibition of the small-scale radial motion yields a quieter zone where the magnetic energy can concentrate.

\subsection{\label{res:scalings}Evolution of diagnostics and asymptotic scalings along the path}

\begin{figure}
\centerline{\includegraphics[width=13.5cm]{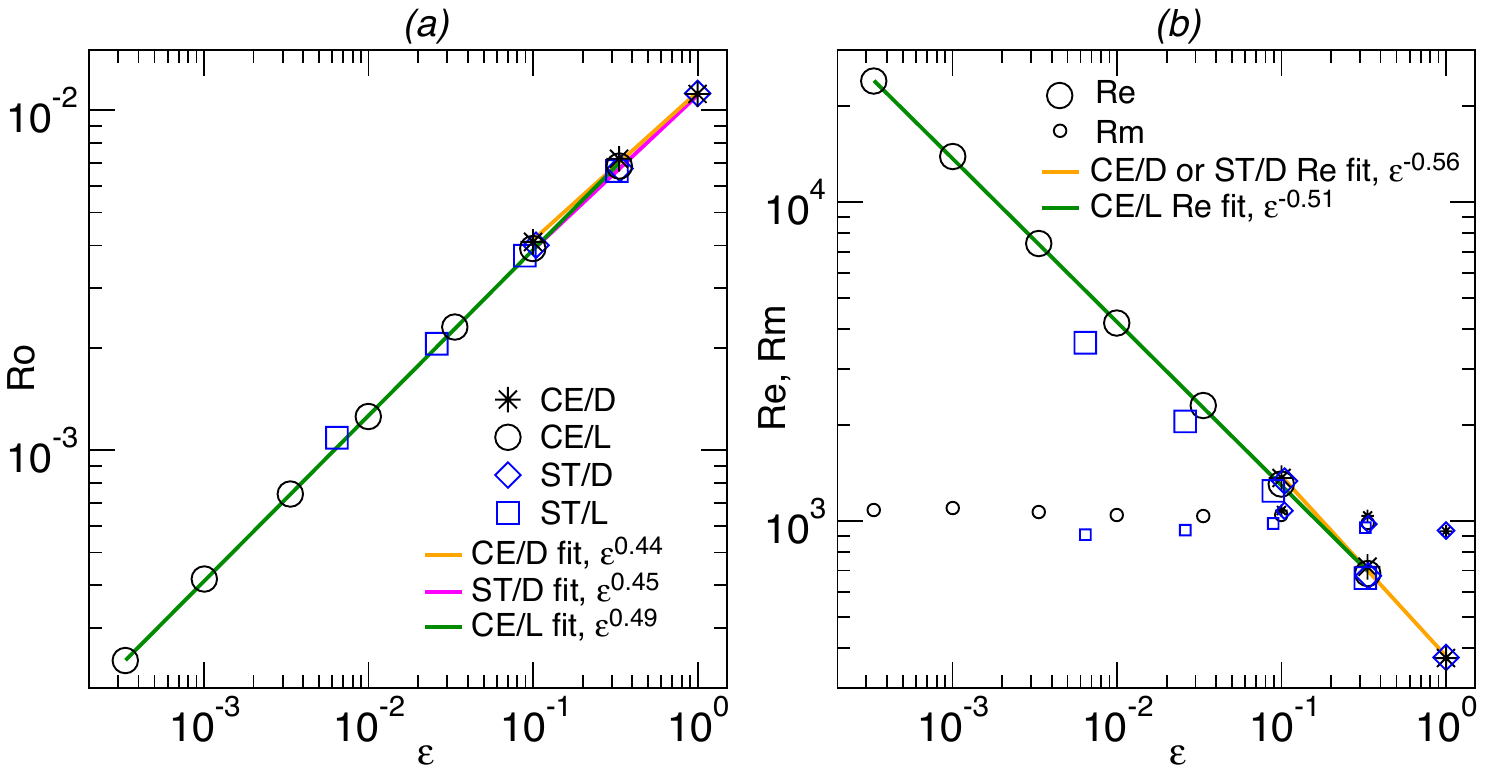}}
\caption{\label{Vscalings} Evolution of the Rossby number $Ro$ ({\it a}), Reynolds $Re$ and magnetic Reynolds $Rm$ numbers ({\it b}) with $\epsilon$.} 
\centerline{\includegraphics[width=13.5cm]{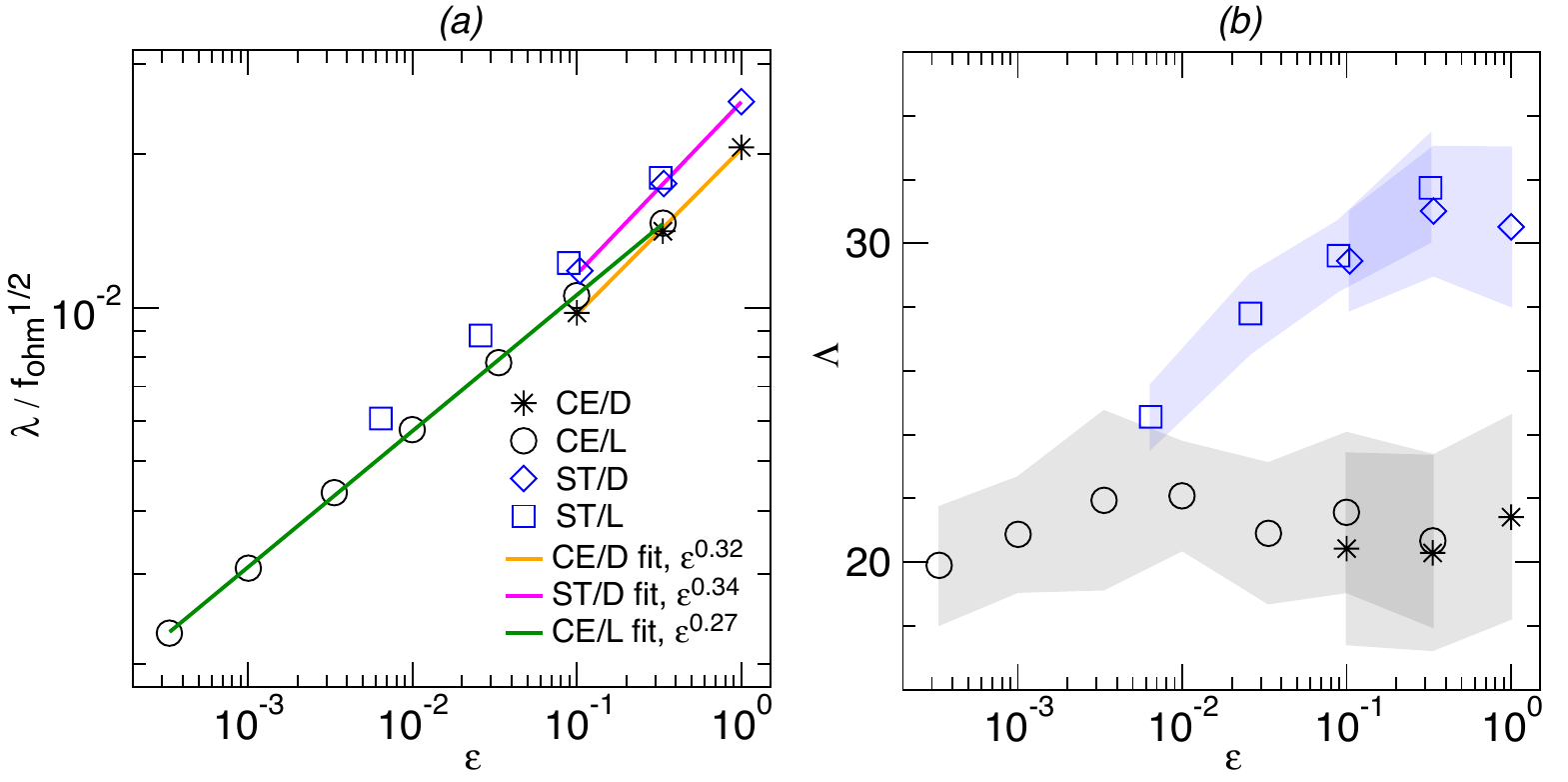}}
\caption{\label{Bscalings} Evolution of the Lehnert number $\lambda$ ({\it a}) and Elsasser number $\Lambda$ ({\it b}) with $\epsilon$. In panel ({\it a}) the Lehnert number is corrected by the factor $f_{\text{ohm}}^{1/2}$ in order to account for the variations of the ohmic dissipation fraction within the scaling theory \citep[e.g.][]{ChristensenAubert2006,Davidson2013}. These variations are significant in the DNS cases but unimportant in the LES cases (see figure \ref{dissip}{\it a}). In the CE-type LES case indeed, the corrected best fit presented in ({\it a}) is closely similar to the uncorrected fit, which yields $\lambda=\lambda_{0} \epsilon^{0.25}$. In panel ({\it b}), shaded regions represent the $\pm 1$ std. dev. of temporal fluctuations relative to the time average.} 
\end{figure}

Figure \ref{Vscalings} presents the evolution of the velocity diagnostics $Ro, Re, Rm$ with $\epsilon$. DNS and LES diagnostics are here again in \rev{broad} agreement. Results obtained with type CE and ST collapse on each other once presented as functions of $\epsilon$, illustrating the convergence between the two types of boundary conditions. CE and ST-type DNS cases respectively follow $Ro=Ro_{0}\epsilon^{0.44\pm 0.02}$,  $Ro=Ro_{1}\epsilon^{0.45\pm 0.01}$, and CE-type LES cases follow $Ro=Ro_{0}\epsilon^{0.49\pm 0.01}$ (figure \ref{Vscalings}{\it a}). These best-fit exponents are in excellent agreement with the D13 prediction 4/9 from equation \rev{(\ref{Roepsilon})} and the path theory prediction 1/2 from scaling (\ref{Roscal}), respectively. This is fully expected since the DNS leaves the spatial structure of the solution free while the LES constrains it in the lateral directions. The slope difference between the two theories is resolvable here thanks to the extremely low scatter of the numerical data and to the 3 decades available in LES cases. The path approach is thus demonstrably advantageous when compared to systematic samplings of the parameter space \citep[e.g.][]{ChristensenAubert2006}. The best-fit exponent 0.46 obtained for $Ro$ in the ST-type LES simulations is significantly below the predicted value 1/2, because the $\epsilon$ values of these simulations are misplaced relative to the values that they should have along the ideal path, as we have seen in section \ref{STpath} and table \ref{table}. It is thus generally not advisable to attempt an interpretation of ST-type LES scaling exponents. Finally, the magnetic Reynolds number (figure \ref{Vscalings}{\it b}) is confirmed to be roughly constant, as already seen in table \ref{table}. 

Figure \ref{Bscalings} presents the evolution of the magnetic field diagnostics $\lambda$ and $\Lambda$ with $\epsilon$. CE and ST-type calculations show more difference than in figure \ref{Vscalings}, but still produce very comparable results. DNS and LES diagnostics are again in good agreement. Within the standard error 0.01 on exponent determination, CE and ST-type DNS cases (figure \ref{Bscalings}{\it a}) are in close agreement with the ohmic fraction-corrected scaling $\lambda/f_{\text{ohm}}^{1/2} \sim \epsilon^{1/3}$ (see equation \rev{\ref{lambdaepsilon}}) expected from D13 and \cite{ChristensenAubert2006}. LES calculations performed in the CE setup yield the corrected Lehnert number scaling $\lambda/f_{\text{ohm}}^{1/2}\sim\epsilon^{0.27\pm0.01}$, and the uncorrected scaling $\lambda=\lambda_{0} \epsilon^{0.25\pm 0.01}$, both also in close agreement with the prediction (\ref{Lescal}) from the path theory. We do not attempt to scale the LES simulations performed in the ST setup, for the same reason as above. The Elsasser number $\Lambda$ is confirmed to be roughly constant (figure \ref{Bscalings}{\it b}) \rev{in CE simulations. The CE-type DNS cases thus appear to invalidate the D13 scaling (\ref{Lambdaepsilon}), but this \revv{is} the consequence of the persistent increase in $f_{\text{ohm}}$ not \revv{being} accounted for in (\ref{Lambdaepsilon}). ST simulations have a higher $\Lambda$ baseline than CE simulations, and the LES cases show a residual decrease with decreasing $\epsilon$}, which can be explained by noting that their rotation rate increases too rapidly with respect to the injected power, as witnessed by the misplaced values of $\epsilon$ relative to the ideal path (table \ref{table}). 

\begin{figure}
\centerline{\includegraphics[width=13.5cm]{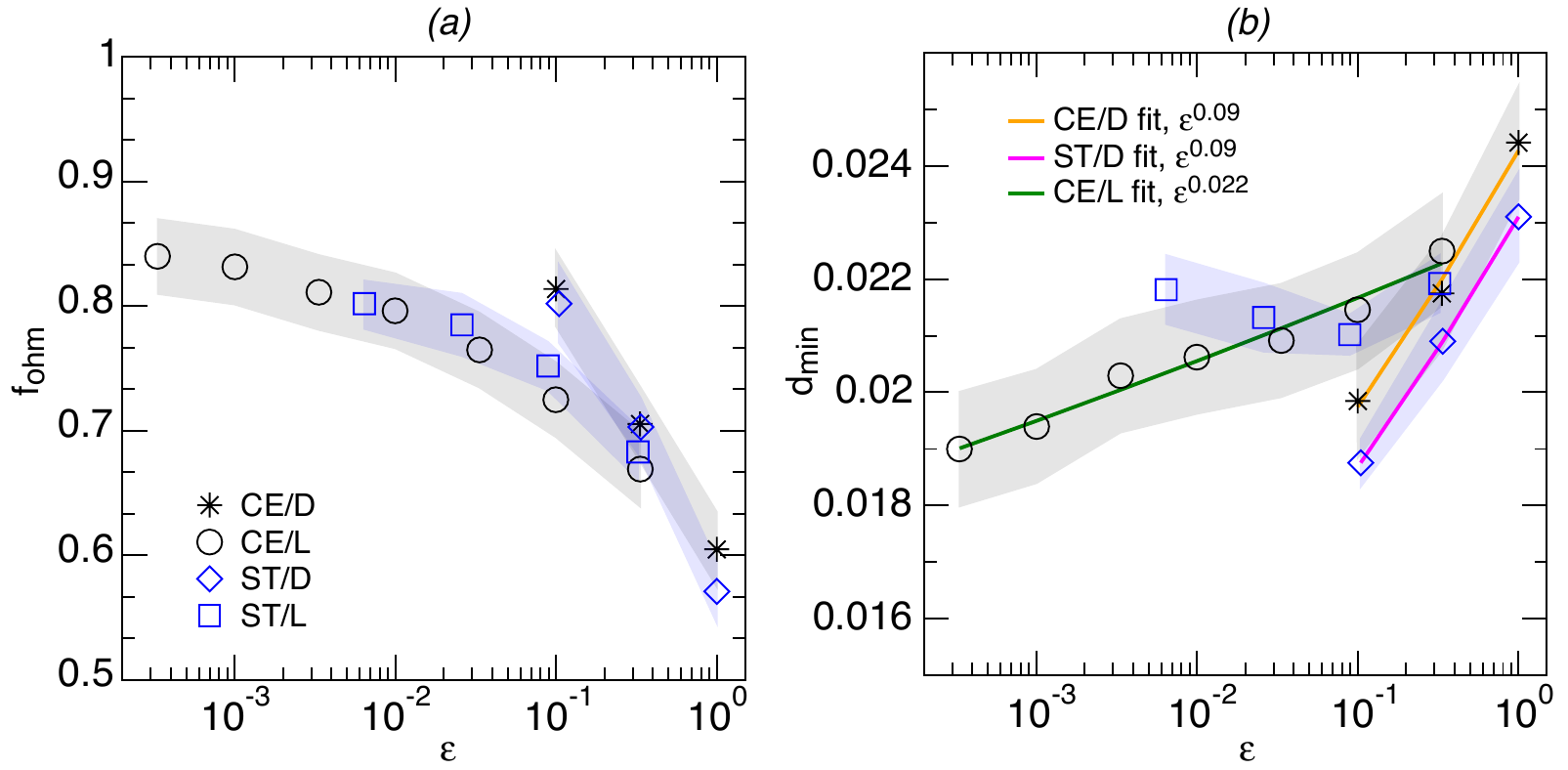}}
\caption{\label{dissip} Evolution of the ohmic dissipation fraction $f_{\text{ohm}}$ ({\it a}) and the magnetic dissipation length $d_{\text{min}}$ ({\it b}) with $\epsilon$. Shaded regions represent the $\pm 1$ std. dev. of temporal fluctuations relative to the time average.} 
\centerline{\includegraphics[width=13.5cm]{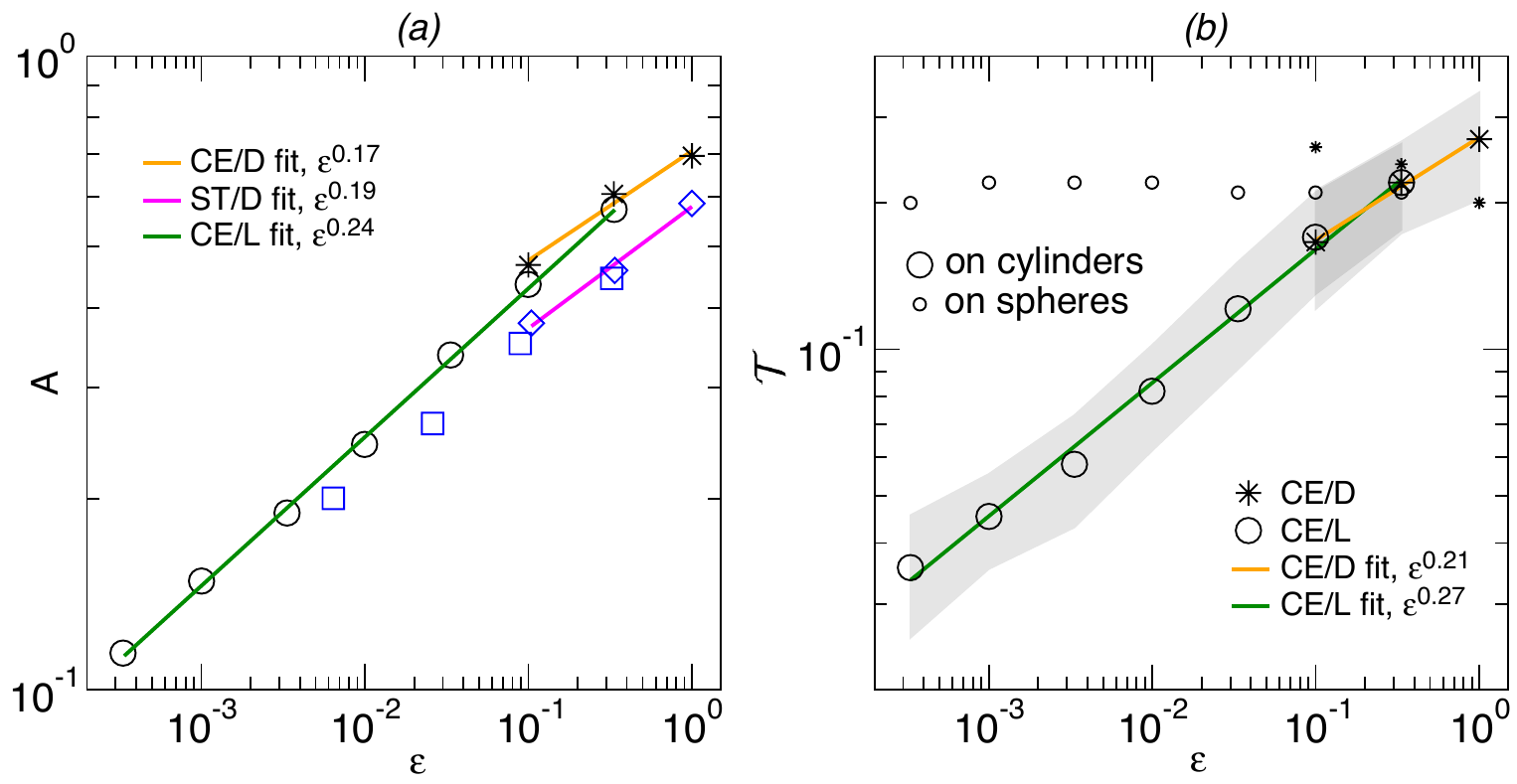}}
\caption{\label{taylor} Evolution of the Alfv\'en number $A$ ({\it a}), and of the level of Taylor constraint enforcement $\mathcal{T}$ ({\it b}) with $\epsilon$. The Alfv\'en number in panel ({\it a}) is the square root of the kinetic to magnetic energy ratio. In panel ({\it b}), the Taylor constraint enforcement level on cylindrical surfaces (large symbols) is as defined in equation (\ref{tayloreq}). For reference, the enforcement level on spherical surfaces (where $\mathcal{T}$ should be unconstrained) is also presented (small symbols). In panel ({\it b}), shaded regions represent the $\pm 1$ std. dev. of temporal fluctuations relative to the time average.}
\end{figure}

We next turn to the analysis of the dissipation diagnostics $f_{\text{ohm}}$ and $d_{\text{min}}$ in our simulations (figure \ref{dissip}). The ohmic dissipation fraction $f_{\text{ohm}}$ (figure \ref{dissip}{\it a}) is largely similar in CE and ST cases. LES simulations produce $f_{\text{ohm}}$ values somewhat below those obtained from DNS simulations performed at similar values of $\epsilon$, an expected consequence of increased viscous losses due to hyperdiffusivity. However, as $\epsilon$ decreases these additional losses are dominated by the increasing ohmic losses, such that LES simulations gradually approach the expected asymptotic behavior $f_{\text{ohm}}=1$. This shows that along the chosen path, the additional viscous losses indeed become asymptotically negligible, thus validating our hyperdiffusive treatment of turbulence. The magnetic dissipation length scale $d_{\text{min}}$ is confirmed to be \rev{weakly variable} within the investigated $\epsilon$ range (note the narrow linear scale used for the ordinate axis of figure \ref{dissip}{\it b}). \rev{A closer analysis shows that} the DNS simulations in ST and CE setups support a decrease $d_{\text{min}} \sim \epsilon^{0.09\pm0.01}$, \rev{matching the dependence $\epsilon^{1/12}$ predicted by the D13 theory (equation \ref{dminepsilon})}. LES simulations in the ST setup do not show a systematic evolution of $d_{\text{min}}$ with $\epsilon$. LES simulations in the CE setup show a weak variation $d_{\text{min}} \sim \epsilon^{\rev{0.022\pm 0.002}}$. \rev{This} can be expected because the LES constrain the length scale to be constant in the two lateral directions, leaving only the radial length scale free to evolve. The discrepancy between DNS and LES, or between the D13 and path theories should remain quite low at Earth's core asymptotic conditions. For $\epsilon=10^{-7}$ we indeed predict $d_{\text{min}} \approx 0.006$ \rev{(corresponding to about 14 km)} if we follow the DNS trend, and $d_{\text{min}} \approx \rev{0.015}$ \rev{(or 34 km)} following the LES trend.  

Finally, we investigate in figure \ref{taylor} the level of enforcement of strong-field dynamo action and of the Taylor constraint in our  simulations. Figure \ref{taylor}({\it a}) shows that at low values of $\epsilon$, the solutions have a kinetic energy much smaller than the magnetic energy, as witnessed by the low values of the squared Alfv\'en number $A^{2}$ measuring their ratio. Together with values of the Elsasser number $\Lambda$ in excess of 1 (figure \ref{Bscalings}{\it b}), this clearly characterises a regime of strong-field dynamo action. The power-law exponents obtained for $A$ in the DNS simulations range between 0.17 and 0.19, in consistency with the results obtained in figures \ref{Vscalings} and \ref{Bscalings}. LES calculations performed in the CE setup yield  $A=A_{0}\epsilon^{\rev{0.24}\pm0.01}$, in close agreement with the prediction (\ref{Ascal}) from the path theory. Analysing the variations of $\mathcal{T}$ in CE-type simulations (figure \ref{taylor}{\it b}), we find that the level of Taylor constraint enforcement increases (i.e. $\mathcal{T}$ decreases) steadily with decreasing $\epsilon$, both in DNS and LES simulations. This again confirms that the hyperdiffusive treatment performed in the LES simulations introduces dynamically negligible additional amounts of the viscous force in asymptotic conditions. Respectively to the time-averaged values of $\mathcal{T}$, we find low instantaneous deviations of $\pm 20 \%$, meaning that the Taylor constraint is enforced at all times in addition to holding everywhere in the fluid domain. As theoretically expected, cylindrical reference surfaces are particular for the Taylor constraint in the sense that the integral Lorentz force remains unconstrained on other surfaces such as spheres. \rev{Qualitatively,} the enforcement of the Taylor constraint links with strong-field dynamo action and the low value of the Alfv\'en number $A=\tau_{A}/\tau_{U}$. This ratio indeed also determines how many torsional oscillation periods $\tau_{A}$ can occur within a convective overturn time $\tau_{U}$ to minimise (by virtue of the Lenz law) the Taylor state deviations that trigger these oscillations. \rev{More quantitatively, the level of Taylor constraint enforcement should scale like the ratio of inertia at the Alfv\'en time scale and magnetic forces at scale $d_{\perp}$, leading to (in dimensional form)
\begin{equation}
\mathcal{T} \sim \rho \mu \dfrac{|\partial\vecu/\partial t|}{|(\nabla\times\vecB)\times\vecB|} \sim \dfrac{d_{\perp}}{\tau_{A}} \rho \mu \dfrac{U}{B^{2}} \sim \dfrac{d_{\perp}}{D} \dfrac{\tau_{U}}{\tau_{A}} \rho \mu \dfrac{U^{2}}{B^{2}} \sim \dfrac{d_{\perp}}{D}A^{2}/A \sim \dfrac{d_{\perp}}{D} A.\label{Tayscal}
\end{equation}}
This suggests that $\mathcal{T}$ should primarily scale with $A$, \rev{as evidenced by the} similarity between the power-law exponents obtained in figures \ref{taylor}({\it a,b}). \rev{We thus propose a scaling law of the form $\mathcal{T}=\mathcal{T}_{0}\, A d_{\perp}/A_{0} (d_{\perp})_{0}$. Using $A_{0}=0.69$, $\mathcal{T}_{0}\approx 0.2$ and $d_{\perp}/(d_{\perp})_{0}=\epsilon^{1/9}$ (equation \ref{mperpepsilon}), at Earth's core conditions i.e. $\epsilon=10^{-7}$ and $A=1.2\te{-2}$ we predict} 
\begin{equation}
\mathcal{T}=\mathcal{O}(10^{-3}),\label{EarthTay}
\end{equation}
such that the Taylor constraint should be strongly but not completely enforced, to leave the possibility for torsional oscillations to exist.

\section{\label{discu}Discussion}
We have defined a path in parameter space connecting classical spherical convective dynamo models to the asymptotic conditions of the Earth's core. This unidimensional path is constrained by the MAC balance and the requirement to preserve the magnetic Reynolds number. Large-scale spatial invariance has been obtained over half the logarithmic distance between models and Earth (figures \ref{spec}-\ref{Bplanforms}). Increasing the rotation rate and power input along the path, we observe direct evidence for a gradual enforcement of the MAC balance (figure \ref{forcebal}), of the Taylor constraint (figure \ref{taylor}), and of strong-field dynamo action with Elsasser numbers in excess of unity  (figure \ref{Bscalings}) and magnetic energy largely dominating the kinetic energy (figure \ref{taylor}). The first half of the path that we have covered is thus devoid of abrupt regime transitions. Furthermore, the model outputs are in excellent agreement with asymptotic, diffusivity-free scalings that accurately predict the Earth's core state. Abrupt transitions are hence also unlikely to happen within the second half of the path. We therefore confirm the asymptotic validity of the D13 theory, and show that dynamo modelling has now advanced to a stage where numerical solutions are sufficiently close to the asymptotic regime. Along the path, the MAC balance, strong-field dynamo action and the Taylor state are furthermore enforced everywhere in the fluid domain, at all instants, and irrespectively of the convective supercriticality, which increases together with the rotational constraint. Our new models thus systematise the results previously obtained in the classical parameter space, where some of these properties were suggested through indirect evidence (see section \ref{intro}), close to the onset of convection (because of the limited rotational constraint), and on the basis of temporal or spatial averages. The way our solutions approach the asymptotic Taylor state by minimising viscosity and inertia is also complementary to an alternative theoretical approach aiming at finding an exact Taylor state, which recently led to numerical solutions in the mean-field framework \citep{Wu2015}. According to our extrapolation on Taylor constraint enforcement in Earth's core (figure \ref{taylor} and \rev{equations \ref{Tayscal}},\ref{EarthTay}), we emphasize that an exact Taylor state is an approximation of Earth's core regime to a similar extent as our numerical solutions, because inertia should be retained for the observed torsional oscillations to exist \citep{Gillet2010}. 

Our solutions at $\epsilon \ll 1$ clearly belong to the inviscid, strong-field, magnetostrophic solution branch conjectured to exist at asymptotic conditions of rapid rotation \citep[e.g.][]{Soward1974,Malkus1975,Roberts1978,Hollerbach1996}. The smooth transition that we observe between classical dynamos and this asymptotic state may be seen as contrasting with the scenario of a catastrophic transition between a weak-field, viscously-dominated solution branch and this strong-field branch \citep[see e.g.][]{Dormy2016}. This apparent contradiction disappears if one considers that classical numerical solutions obtained over the past decade have been sometimes incorrectly attributed to the weak-field, viscous branch. It is true that such solutions, which serve as a starting point for our path, are indeed not strong-field since their kinetic and magnetic energies are comparable, at the exception of a few cases obtained at low supercriticality \citep[i.e. by minimising the kinetic energy,][]{Takahashi2012,Dormy2016}. But they are not viscous either, as their force balance is already a well-satisfied MAC equilibrium (figure \ref{forcebal}) with sizeable residual contributions of inertial and viscous forces. These secondary contributions gradually disappear as we progress along the path, and the separation between kinetic and magnetic energy gradually increases. In summary, the branch on which most classical dynamos reside does not belong to either of the historical regimes that we alluded to above, but this branch gradually morphs into the strong-field branch as we reach the rapid rotation limit. The weak-field branch is nonexistent far above the onset of convection \citep[e.g.][]{Roberts1978}, rationalising the fact that we did not observe bistability in our strongly supercritical simulations.

Large-eddy simulations are key to the numerical feasibility of asymptotic, rapidly rotating convective dynamos. They are made possible because of the \rev{large-scale} spatial invariance conjectured \rev{to hold at a constant level $Rm$ of magnetic turbulence, and broadly confirmed by} direct numerical simulations. Here we have used an hyperdiffusive treatment that does not perturb the large-scale MAC balance structure, and asymptotically introduces a negligible additional amount of the viscous force (figures \ref{forcebal},\ref{taylor}) and of viscous dissipation (figure \ref{dissip}), leaving ohmic losses as the dominant source of dissipation. \rev{Compared to direct numerical simulations, the most accurate large-eddy simulation results (figures \ref{specdnsvsles}-\ref{Eprofiles}) have been obtained by using fixed mass anomaly fluxes at the boundaries, yielding a better control on the convective power and a transport dominated by large scales. Furthermore,} extreme calculations are also made possible by using boundary conditions that dispense of viscous and density anomaly boundary layers. \rev{Their scaling properties} have been demonstrated to be largely identical to calculations where these boundary layers are present (e.g. figure \ref{Eprofiles}). In particular, we have not observed evidence of active boundary layers, such as recently proposed by \cite{Stellmach2014}. Our interpretation is that the configuration used by these authors mostly applies to the tangent cylinder of a spherical shell, while most of the energy transfer in our simulations occurs elsewhere in low-latitude regions \citep[][]{Yadav2016}. We also obtain zonal flows of similar amplitude regardless of mechanical boundary conditions, another confirmation that they are thermal-wind limited in dipole-dominated spherical convective dynamos \citep{Aubert2005,Yadav2013FS}, and that possible residual effects of the boundary viscous drag \citep{Livermore2016} are not present in our calculations.

As mentioned above, diffusivity-free, power-driven scaling laws can be derived and numerically validated for the velocity, magnetic and density anomaly fields amplitude along the path (equations \ref{Roscal},\ref{Lescal},\ref{Cscaling}, figures \ref{Tplanforms},\ref{Vscalings},\ref{Bscalings}) and yield predictions in striking agreement with geophysical estimates. Crucial to this success and also to the consistency of the path theory is that our estimate of Earth's core flux-based, modified Rayleigh number $Ra^{*}_{F}$ (equation \ref{RaEarth}) is probably about correct. Another important point is that the path theory underlying these scalings is a spatially-invariant approximation of the D13 theory, such that there exists a few differences in scaling exponents between the path and D13 theories \rev{(see section \ref{CEpath})}. These differences are resolvable in our numerical dataset (for instance in figures \ref{Vscalings},\ref{Bscalings}), with the direct numerical simulations in full agreement with D13 and the large-eddy simulations supporting the path theory. For theoretical purposes, the D13 theory should be used, but for most geophysical purposes the path theory suffices because the differences between the two theories are minimal at Earth's core conditions. As an illustration, predictions for the velocity, magnetic field amplitudes, and magnetic dissipation length differ by only a factor 2 to 4 at $\epsilon=10^{-7}$. Still, \rev{throughout the path the D13 theory predicts a somewhat larger (roughly a factor 6, equation \ref{mperpepsilon}) decrease for the large scale $d_{\perp}$. This decrease is not observed in direct numerical simulations covering the first decade of the path (figure \ref{forcebal}) because the ohmic fraction $f_{\text{ohm}}$ influencing $d_{\perp}$ (equation \ref{dperp}) still varies significantly (figure \ref{dissip}{\it a}) and compensates for the expected variation. These results outline the limits of our present approach, and suggest that more extreme direct numerical simulations or large-eddy simulations at higher resolution will still be required to fully capture the effects of the residual variations of $d_{\perp}$ and $d_{\text{min}}$ along the path.} \revv{Such simulations will also be needed to clarify the control of inertia on the occurrence of magnetic polarity reversals \citep[e.g.][]{ChristensenAubert2006,Sheyko2016}, a topic that we have kept outside the scope of the present study.}

There has been a long-standing debate on the relevance of classical numerical solutions to the structural and mechanistic description of the geodynamo \citep[see e.g. a discussion in][]{Christensen2015}, that our results should contribute to settle. Indeed we have shown that classical models at the starting point of our path have a large-scale spatial structure which is essentially invariant as we progress towards asymptotic conditions, and a reasonably well enforced MAC balance. This suffices to produce solutions with statics and kinematics that compare favourably to the observable geomagnetic field \citep{Christensen2010,Aubert2013b}. However, classical models fall short of fully accounting for the dynamics because the amplitude of the magnetic force is incorrect relatively to core inertia (they are not in a strong-field regime). Further down the path, the strong-field models that we have calculated open a window on the short-timescale magnetic dynamics that the classical models fail to resolve. One topic of interest concerns magnetohydrodynamic waves such as torsional oscillations and MAC waves. These can be clearly exhibited in contexts where the magnetic field is forced \citep[see e.g.][]{Braginsky1993,Braginsky1995,Buffett2014} but tend to become quite subtle in free, self-sustained dynamo systems \citep{Wicht2010,Teed2015} because of the Lenz law effects that follow from strong-field dynamo action. Our new models may help to assess the levels at which these waves settle in free environments, and their possible geophysical signatures.

By construction, our large-eddy simulations do not investigate the turbulence underlying the large-scale MAC system that we have exhibited. In this respect, direct numerical simulations and laboratory experiments remain essential \citep[see in particular][]{Aurnou2015,Nataf2015}. Since turbulence occurs below the magnetic dissipation length scale, one could conjecture that it may be magnetically unconstrained. It would thus be interesting to assess how much the small-scale system has then in common with quasi-gestrophic, two-dimensional turbulence. In that sense, the ideas brought forward within the context of a multi-scale dynamo model \citep{Calkins2015} could help to formulate a unified model handling both the large (MAC) and small (geostrophic) scales. Though it has been demonstrated to be fairly accurate, our large-scale approximation could also certainly be made better through progress on the characterisation and modelling of turbulent fluxes between large and small scales.

\begin{acknowledgements}
We thank Ulrich Christensen and two anonymous referees for insightful comments that helped to enhance the manuscript. JA acknowledges support from French {\it Programme National de Planétologie} (PNP) of CNRS/INSU. This work was granted access to the HPC resources of S-CAPAD, IPGP, France, and to the HPC resources of TGCC, CINES and IDRIS under the allocation 2016-042122 made by GENCI. This is IPGP contribution \rev{3781}. 
\end{acknowledgements}

\bibliographystyle{jfm}
\bibliography{Biblio}

\end{document}